\documentclass{article}

\usepackage{arxiv}

\usepackage[utf8]{inputenc} 
\usepackage[T1]{fontenc}    
\usepackage{hyperref}       
\usepackage{url}            
\usepackage{booktabs}       
\usepackage{amsfonts}       
\usepackage{nicefrac}       
\usepackage{microtype}      
\usepackage{lipsum}
\usepackage{fancyhdr}       
\usepackage{graphicx}       
\graphicspath{{media/}}     
\usepackage[square,sort,comma,numbers]{natbib}
\usepackage{adjustbox}
\usepackage{verbatim}
\usepackage{bm}
\usepackage{tabularx}			
\usepackage{longtable}

\usepackage{nicematrix}
\usepackage{array}
\usepackage{makecell}
\usepackage{tabularray}
\usepackage{array,booktabs}%
\usepackage{graphics}

\usepackage{graphicx}
\usepackage{newtxtext}
\usepackage{newtxmath}
\usepackage{natbib}
\usepackage{hyperref}
\usepackage{subcaption}	
\usepackage{engord}

\usepackage{lineno}
\usepackage{multirow}

\title{Solution multiplicity and effects of data and eddy viscosity on Navier-Stokes solutions inferred by physics-informed neural networks
}

\author{
  Zhicheng Wang \\ 
  Laboratory of Ocean Energy Utilization of Ministry of Education, \\
  School of Energy and Power Engineering, \\ 
  Dalian University of Technology. \\
 \texttt{zhicheng\_wang@dlut.edu.cn}
  \And
  Xuhui Meng \\
  Institute of Interdisciplinary Research for Mathematics and Applied Science, \\
  School of Mathematics and Statistics,\\
  Huazhong University of Science and Technology
   \And
  Xiamo Jiang \\
  School of Energy and Power Engineering, \\
  State Key Lab of Structural Analysis, Optimization and CAE Software for Industrial Equipment, \\
  Research Institute for Carbon Neutrality,\\
  Dalian University of Technology \\
  \And
  Hui Xiang \\
  Research Institute for Carbon Neutrality,\\
  Dalian University of Technology \\
  \And
   George Karniadakis \\
  Division of Applied Mathematics and School of \\
  Providence, RI\\
  \texttt{george\_karniadakis@brown.edu} \\
}

\begin{document}
\maketitle

\begin{abstract}
Physics-informed neural networks (PINNs) have emerged as a new simulation paradigm for fluid flows and are especially effective for inverse and hybrid problems. However, vanilla PINNs often fail in forward problems, especially at high Reynolds (Re) number flows. Herein, we study systematically the classical lid-driven cavity flow at $Re=2,000$, $3,000$ and $5,000$. We observe that vanilla PINNs obtain two classes of solutions, one class that agrees with direct numerical simulations (DNS), and another that is an unstable solution to the Navier-Stokes equations and not physically realizable. We attribute this solution multiplicity to singularities and unbounded vorticity, and we propose regularization methods that restore a unique solution within 1\% difference from the DNS solution. In particular, we introduce a parameterized entropy-viscosity method as artificial eddy viscosity and identify suitable parameters that drive the PINNs solution towards the DNS solution. Furthermore, we solve the inverse problem by subsampling the DNS solution, and identify a new eddy viscosity distribution that leads to velocity and pressure fields almost identical to their DNS counterparts. Surprisingly, a single measurement at a random point suffices to obtain a unique PINNs DNS-like solution even without artificial viscosity, which suggests possible pathways in simulating high Reynolds number turbulent flows using vanilla PINNs.
\end{abstract}


\section{Introduction}
\label{sec:intro}
The incompressible viscous Newtonian fluid flow driven by a lid at constant speed enclosed in a square cavity is the most well studied prototype flow, owing to its simplicity of the geometry, the mesh generation and the implementation of boundary conditions. In two-dimensions (2D), the flow remains steady up to about Reynolds number $Re \approx 8,000$\citep{AUTERI20021}, while in 3D it becomes unsteady at $Re \approx 800$ \citep{theofilis_duck_owen_2004}. Despite the plethora of studies, the cavity flow may present intriguing phenomena, e.g., that the steady solution at a relatively high Reynolds number ($Re$) may not be unique \citep{Cavity_ARFM_2000,Kuhlmann2018TheLC}. This is consistent with theoretical work \citep{yudovich_1967,galdi2011introduction} on the Navier-Stokes equations that states that  there is a unique steady state solution provided we have a problem with large viscosity or small `data', i.e., small Reynolds number. However, when the viscosity gets smaller (or the boundary data gets larger) one would expect some kind of bifurcation, which translates into the co-existence of multiple stable and unstable states that cannot be captured by direct numerical simulation (DNS).

In the three-dimensional visualization experiment by \citep{Cavity_PoF_1991}, it was found  that if the lid suddenly decelerates the flow from $Re=2,000$ to $Re=500$, the original primary state may or may not be recovered. In the three-dimensional numerical simulation, the authors in \citep{kuhlmann_wanschura_rath_1997} reported that there exist two solutions stable to 2D disturbances in the cavity flow with a pair of opposite walls moving at the same speed but in opposite directions.  Using different numerical methods, 
several studies  \citep{Albensoeder2001MultiplicityOS,ARUMUGAPERUMAL20113711,PRASAD2016297} found multiple 2D steady states in the rectangular double-lid-driven cavities. More recently, it was found that there also exist multiple solutions in the lid-driven right-angled isosceles triangular \citep{an_bergada_mellibovsky_2019} as well as for the cavity with an arc-shaped wall \citep{Cavity_arc_2021}. Nonetheless, to the best of the authors' knowledge, no study has shown that multiple solutions are possible in the 2D single-lid-driven rectangular cavity, including any of the numerical simulation studies by using conventional methods.   

Physics-Informed Neural Networks (PINNs) have been developed as an alternative method to solve partial differential equations (PDEs), especially in the presence of data \citep{RAISSI2019686}. Compared to the conventional methods, the PINNs model is a mesh-free method that can easily handle complex geometries \citep{Karniadakis2021PhysicsinformedML}. In addition, it can predict the solution accurately in an arbitrary location within the domain by leveraging the interpolation capability of neural network and physics. This approach has been applied to tackle ill-posed fluid flow problems, where boundary conditions or equation parameters are not well defined \citep{Cai2021PhysicsinformedNN, JIN2021109951}. Specifically, it has been recently applied to address the fundamental fluid mechanics problem whether there exist finite time blow-up solutions for the 2D Boussinesq and the 3D Euler equations\citep{wang2023asymptotic}. 

Despite the success of PINNs in predicting general fluid flows \citep{JIN2021109951}, there still remains significant challenges in simulating fluid flows at a higher $Re$. In particular, for the cavity flow at $Re \ge 1000$, in the scenario that no labeled data (i.e., measurements) are available, no public report can be found about PINNs solution in agreement with that obtained by accurate numerical simulation, except our own work at $Re=1,000$ \citep{He2023}, where we proposed the entropy viscosity method (EVM) to improve the PINNs inference. In the current paper, we will employ a parametric EVM as well as discover a neural network that implicitly represents eddy viscosity in order to simulate  the cavity flow at $Re = 2,000$, $Re=3,000$ and $Re=5,000$. Interestingly, we find that the original PINNs obtain multiple solutions of the Navier-Stokes equations, which have not been captured before with the traditional numerical methods of computational fluid dynamics.  Hence, PINNs is a new paradigm complementary to DNS that captures only the most energetic solution from a multitude of possible solutions to the Navier-Stokes equations.

\section{Entropy-viscosity for PINNs}

We employ PINNs to solve the following 2D steady Navier-Stokes equations with appropriate boundary conditions,
\begin{equation}\label{NSE1}
    \mathbf{u}\cdot \nabla \mathbf{u}=-\nabla p + \frac{1}{Re}\nabla^2 \mathbf{u}, \quad \text{in}\, \Omega, 
\end{equation}
\begin{equation}\label{NSE2}
    \nabla \cdot \mathbf{u}=0, \quad \text{in}\, \Omega, 
\end{equation}
\begin{equation}\label{BC1}
    \mathbf{u}=\mathbf{u}_{\Gamma},\quad \text{on} \,\partial \Omega_{\Gamma},
\end{equation}
\begin{equation}\label{BC2}
    \frac{\partial \mathbf{u}}{\partial \mathbf{n}}=\mathbf{g}_{\Pi},\quad \text{on} \,\partial \Omega_{\Pi},
\end{equation}
where, $\mathbf{u}(x,y)=[u,v]^{T}$ is the velocity vector, and $p$ is the pressure; $\Omega$ denotes the domain, $\partial \Omega_{\Gamma}$ and $\partial \Omega_{\Pi}$ are the Dirichlet and Neumann boundaries, respectively. $Re=\frac{\mathbf{U}_{\infty}L}{\nu}$ is the flow Reynolds number, where $\mathbf{U}_{\infty}$ and $L$ are the characteristic velocity and length, respectively, and $\nu$ is the kinetic viscosity.  

Following the PINNs framework described in \citep{Cai2021PhysicsinformedNN,JIN2021109951}, a fully connected feed-forward neural network (FNN) is used to approximate the solution of Eq. \ref{NSE1} and Eq. \ref{NSE2} subjected to the boundary conditions Eq. \ref{BC1} and Eq. \ref{BC2}. The neural network consisting of multiple hidden layers takes the space coordinates $\mathbf{x}$ as its input and $\mathbf{u},\,p,\, r$ as the output. In particular, the hidden variables $Y$ in $k^{th}$ layer can be calculated as follows,
\begin{equation}
Y^{k} = \sigma (W^{k}Y^{k-1} + B^{k}),
\end{equation}
where $W$ and $B$ are the weight matrix and bias vectors, respectively, which will be updated iteratively during training.  Note that we use the hyperbolic tangent activation function, namely $\sigma(\cdot)=\tanh{(\cdot)}$.
Specifically, solving the governing equations \ref{NSE1}-\ref{BC2} is equivalent to minimizing the mean squared error (MSE) of the following loss functions,
\begin{equation}\label{eq:loss1}
  e_1=u\frac{\partial u}{\partial x}+v\frac{\partial u}{\partial y}+\frac{\partial p}{\partial x}-(\frac{1}{Re}+\nu_E)(\frac{\partial^2 u}{\partial x^2}+\frac{\partial^2 u}{\partial y^2}),  
\end{equation}
\begin{equation}\label{eq:loss2}
  e_2=u\frac{\partial v}{\partial x}+v\frac{\partial v}{\partial y}+\frac{\partial p}{\partial y}-(\frac{1}{Re}+\nu_E)(\frac{\partial^2 v}{\partial x^2}+\frac{\partial^2 v}{\partial y^2}),  
\end{equation}
~%
\begin{equation}
e_3 = \frac{\partial u}{\partial x}+\frac{\partial v}{\partial y},
\end{equation}
where $\nu_E$ is the eddy viscosity that will be determined during training. Note that $\nu_E$ is a scalar, whose construction is adapted from the entropy viscosity method \citep{GuermondEVMJCP,wang2019jfm} for numerical stabilization in the flow simulation at a high $Re$. Depending on the scenarios whether some velocity or pressure measurements (labeled data) are  available for training, we develop two approaches for obtaining $\nu_E$ in this paper, as follows:
(1) \emph{neural network model}; and (2) {\emph{parameterized model}}.  
We will employ the first approach in the scenario where a small amount of labeled data is available, while the second one is more suitable in the case of no labeled data. Specifically, in the second approach, $\nu_E$ can be computed from: 
\begin{equation}\label{eq:nu_e}
  \nu_E=\min(\beta \nu , \alpha \frac{|r| L^2}{U_{\infty}^2}),  
\end{equation}
where $r$ is the predicted entropy residual computed as,
\begin{equation}\label{eq:ev}
    r = (u-u_m)e_1+(v-v_m)e_2.
\end{equation}
We note that we insert Eq.\ref{eq:ev} into the neural network losses as follows as residual in the form,
\begin{equation} \label{EV1}
e_4=(u-u_m)e_1+(v-v_m)e_2-r.
\end{equation}
Note that Eq. \ref{EV1} produces non-zero residual on the wall boundary, which is different from the entropy viscosity in the numerical method developed in \citep{wang2019jfm}. We found that $\nu_E > 0$ in the vicinity of the boundary can give riser to better prediction of PINNs. Moreover, in Eq. \ref{eq:nu_e}, $\alpha$ and $\beta$ are two tunable hyper parameters, whose values may affect the PINNs inference greatly; $\alpha$ and $\beta$ can be either constant or descending throughout the training. The latter scenario was found from the inverse problem where both $\alpha$ and $\beta$ were learned from the labeled data, as shown in Fig. S2 in the supplementary information. Here,  $u_m$ and $v_m$ are two constants representing the global mean values of $u$ and $v$ within $\Omega$, respectively; $u_m=0.5 U_{\infty}, and \, v_m=0.5 U_{\infty}$ are used in this paper unless otherwise stated.

With all the components of the loss functions defined, the problem set up is as follows:
\begin{equation}\label{loss_main}
    \arg \min L=L_b+L_e + L_s,
\end{equation}
where, 
\begin{equation}
L_b=\lambda_b\big(\sum_{n=1}^{N_\Gamma}|\mathbf{u}(\mathbf{x}_n)-\mathbf{u}_{\Gamma}(\mathbf{x}_n)|^2+\sum_{n=1}^{N_{\Pi}}|\frac{\partial \mathbf{u}(\mathbf{x}_n)}{\partial \mathbf{n}}-\mathbf{g}_{\Pi}(\mathbf{x}_n)|^2 \big),
\end{equation}
\begin{equation}    L_e=\sum_{i=1}^3\big(\lambda_i\sum_{n=1}^{N_e}|e_i(\mathbf{x}_n)|^2\big),
\end{equation}
\begin{equation}    L_s=\lambda_s \sum_{n=1}^{N_e}|e_4(\mathbf{x}_n)|^2,
\end{equation}
where $L_b$,$L_e$ and $L_s$ represent the loss on the boundary, the governing equation and the entropy residual, respectively. In addition, $\lambda_b=10$, $\lambda_e=1$ and $\lambda_s=0.1$ are used in this paper unless stated otherwise. $N_e$, $N_{\Gamma}$ and $N_{\Pi}$ are the number of sampling points in the domain $\Omega$, Dirichlet boundary $\partial \Omega_{\Gamma}$ and $\partial \Omega_{\Pi}$, respectively.
Furthermore, the parameters of the neural networks ($W$ and $B$) are initialized using the Xavier scheme, and the Adam optimizer is employed in the entire training.


\section{Multiple solutions inferred by PINNs} 

We use the code NSFnet \citep{JIN2021109951}, which we have modified to
include the entropy viscosity (ev) term as shown in the equations above. The simulation domain is the unit square of dimension $[0,1]\times [0,1]$. To avoid strong singularities at the two corners where there is boundary condition discontinuity, we  re-formulate the lid-driven boundary condition as follows,
\begin{align}
u(x,y)= & 1-\frac{\cosh{(C_0(x-0.5))}}{\cosh{(0.5C_0)}},  \\
v(x,y)= & 0,
\end{align}
where $\, x \in [0,1], \,y=1,$ $C_0=50$.

\begin{table}
\caption{ Case names, corresponding parameters and relative percent errors (RPE, $\epsilon$). $N_D$ is the number of labeled data used in the training. Note that the error is defined as $\epsilon_{\phi}=\frac{\| \hat{\phi}-\phi\|_2}{ \|\phi\|_2} \times 100$, where $\phi$ denotes the reference data and $\hat{\phi}$ is the PINNs predicted value, both being computed on the uniform $256\times 256$ mesh. $|\nabla p|$ instead of $p$ is employed to calculate the RPE in pressure field. Hidden neurons $6\times 80$ indicates that the network consists of 6 hidden layers, 80 neurons in each layer. Hidden neurons $4 \times 120 + 4 \times 40$ indicates that $4 \times 120$ neurons are for variables $u,\,v,\,p$, and $4 \times 40$ neurons are for variables $r$. The cases in this table are selected for supporting the main context of this paper. Please refer to Table S1 and Table S2 in the supplementary material for more details.}
\begin{center}
\begin{adjustbox}{width=0.9\textwidth}
\begin{NiceTabular}{|c|c|c|c|c|c|c|c|c|c|}
\toprule
\textcolor{blue}{\textbf{Case}}
& PINNs & Eddy Viscosity Model & Hidden Neurons& $Re$ & $N_D$
&\multicolumn{3}{|c|}{\makecell[c]{$\epsilon_u\quad\, \epsilon_v \quad\, \epsilon_p$}} \\
\midrule
\textcolor{blue}{\textbf{A}} & NSFnet & none & $6 \times 80$ & $2,000$& 0
&\multicolumn{3}{|c|}{\makecell[c]{89.0\, 88.6\, 94.0}} \\
\hline
\textcolor{blue}{\textbf{B}} & ev-NSFnet & \emph{single-network parameterized} & $6 \times 80$  & $2,000$& 0
&\multicolumn{3}{|c|}{\makecell[c]{2.1\, 2.2\, 4.8}} \\
\hline
\textcolor{blue}{\textbf{C}} & ev-NSFnet & \emph{single-network parameterized} & $6 \times 80$  & $3,000$& 0
&\multicolumn{3}{|c|}{\makecell[c]{4.7\, 4.8\, 9.6}} \\
\hline
\textcolor{blue}{\textbf{D}} & NSFnet & none & $6 \times 80$  & $2,000$& 1
&\multicolumn{3}{|c|}{\makecell[c]{0.9\, 0.9\, 1.4}} \\
\hline
\textcolor{blue}{\textbf{E}} & ev-NSFnet & single-network parameterized & $6 \times 80$ & $2,000$& 5
&\multicolumn{3}{|c|}{\makecell[c]{0.5\, 0.5\, 2.6}} \\
\hline
\textcolor{blue}{\textbf{F}} & ev-NSFnet & neural network & $6 \times 80$  & $2,000$& 100
&\multicolumn{3}{|c|}{\makecell[c]{0.3\, 0.3\, 0.7}} \\
\hline
\textcolor{blue}{\textbf{G}} & ev-NSFnet & \emph{two-network parameterized} & $4 \times 120 + 4 \times 40$ & $3,000$& 0
&\multicolumn{3}{|c|}{\makecell[c]{2.8\, 2.8\, 4.6}} \\
\hline
\textcolor{blue}{\textbf{H}} & ev-NSFnet & \emph{two-network parameterized} & $6 \times 80 + 4 \times 40$  & $5,000$& 0
&\multicolumn{3}{|c|}{\makecell[c]{5.4\, 5.4\, 9.3}} \\

\bottomrule
\end{NiceTabular}
\end{adjustbox}
\end{center}
\label{table_Re2K}
\end{table}
\begin{figure}
\centering
\includegraphics[width=0.95\textwidth]{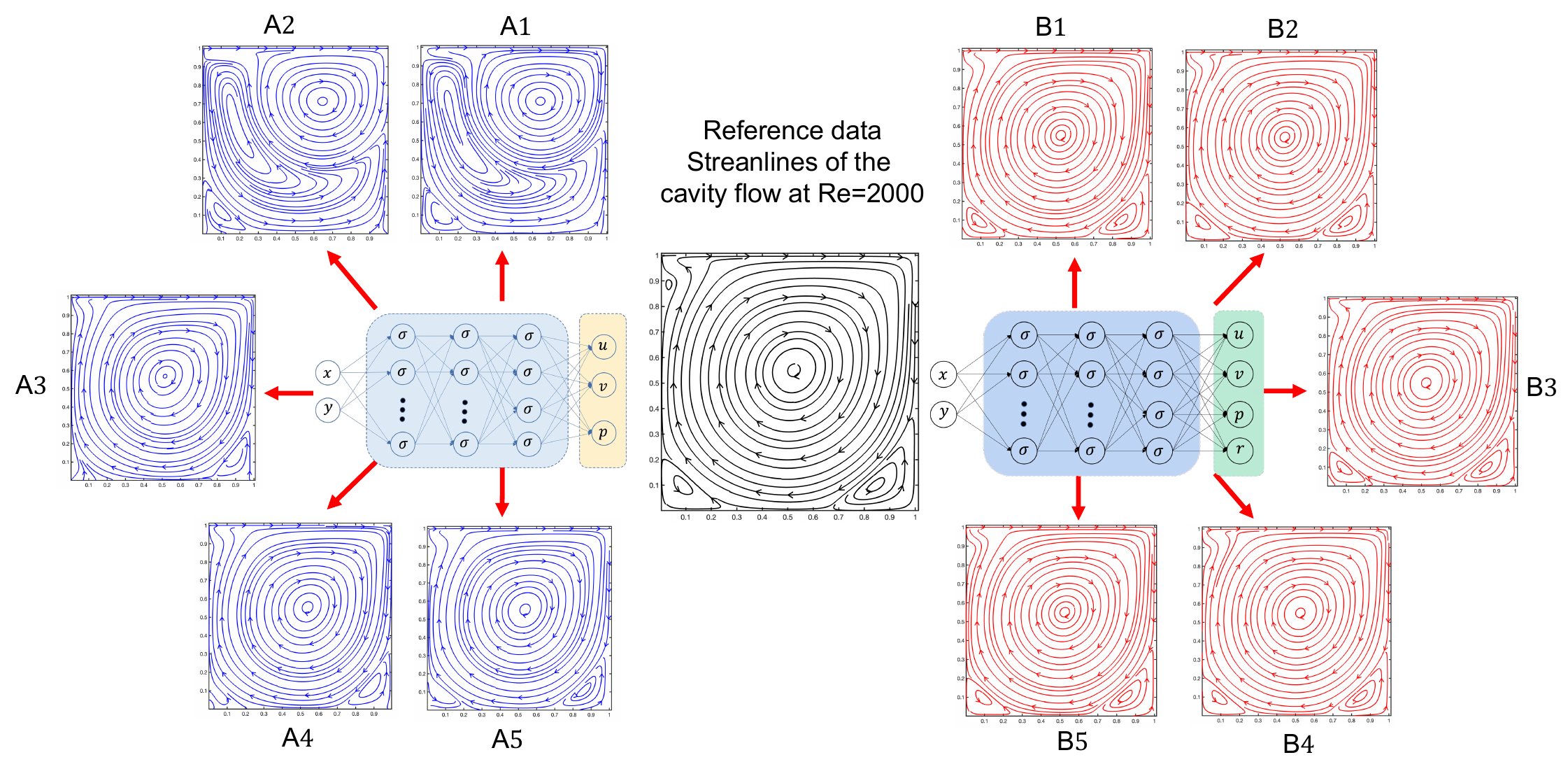}
\caption{\textbf{A plurality of stable and unstable solutions of the 2D cavity flow at } $\boldsymbol{Re=2,000}$. Shown are inference streamlines by the original NSFnet and the entropy-viscosity regularized ev-NSFnet. The five flow plots on each side of the DNS (Reference) solution are obtained based on five independent initialization values of PINNs using the same hyper-parameters. A1 and A2 represent a new flow type that is not captured by DNS or by ev-NSFnet. Adding eddy viscosity to the Navier-Stokes leads to a stable solution, similar with what is captured by DNS. The error of each case in A1-A5 and B1-B5 is given in Table S2 of the supplementary materials.  The average error in the velocity field for B1-B5 is less than 4\%. The `new' type streamlines captured by the original NSFnet and the type captured by ev-NSFnet as well as the corresponding evolution with training can be seen from Movie S1 and S2 in the supplementary material. 
} 
\label{fig:ev_nsf-net}
\end{figure}

Unlike the lid-driven cavity flow at low Reynolds number, at $Re=2,000$
we obtained multiple PINNs solutions, depending on the initialization of the neural networks, namely the weights and the biases. Typical results are shown in figure \ref{fig:ev_nsf-net}. Specifically, 
the left part of figure \ref{fig:ev_nsf-net} demonstrates the capability of PINNs to obtain the multiple stable or unstable solutions of the 2D cavity flow, when no labeled data is used. We observe that the original NSFnet obtains 5 different solutions in 5 independent trains with exactly the same parameters and epochs. More interestingly, the five solutions can be divided into two classes: \emph{class 1}, consisting of solutions A1 and A2; \emph{class 2}, consisting of solutions A3-A5. As shown by the streamlines, in \emph{class 1} there are two large vortices formed, with a  third small vortex in the left-bottom corner, while the three vortices are near symmetric with respect to the counter diagonal of the computational domian. The streamline pattern of \emph{class 2} is similar to the reference data (obtained by direct numerical simulation (DNS) and the spectral element code Nektar \citep{GK_CFDbook}): a large vortex is located in the domain center, and three small vortices develop in the left-top, left-bottom and right-bottom corners. Compared to the reference solution, the relative percentage error (RPE) of the solution in \emph{class 1} is greater than 90\%, while the error in \emph{class 2} is less than 40\%.
However, with the help of entropy viscosity in training, the PINNs optimizer can avoid being stuck at a local minimum leading to the solution \emph{class 1} type and instead  capturing the `correct' solution of \emph{class 2} type, as shown on the right part of figure \ref{fig:ev_nsf-net}. Moreover, the entropy viscosity improves the accuracy of PINNs inference notably. Here the ev-NSFnet solutions obtained in the 5 independent runs all agree with the reference solution very well, with RPE less than 4\%, indicating that the entropy viscosity can lead the PINNs to find the `correct' solution. The detailed RPE value of each case shown in figure \ref{fig:ev_nsf-net} can be found in Table 1 of the supplementary material.

\begin{figure}
\centering
  \begin{subfigure}{\textwidth}
\includegraphics[width=0.32\textwidth,trim=0 145 0 145,clip]{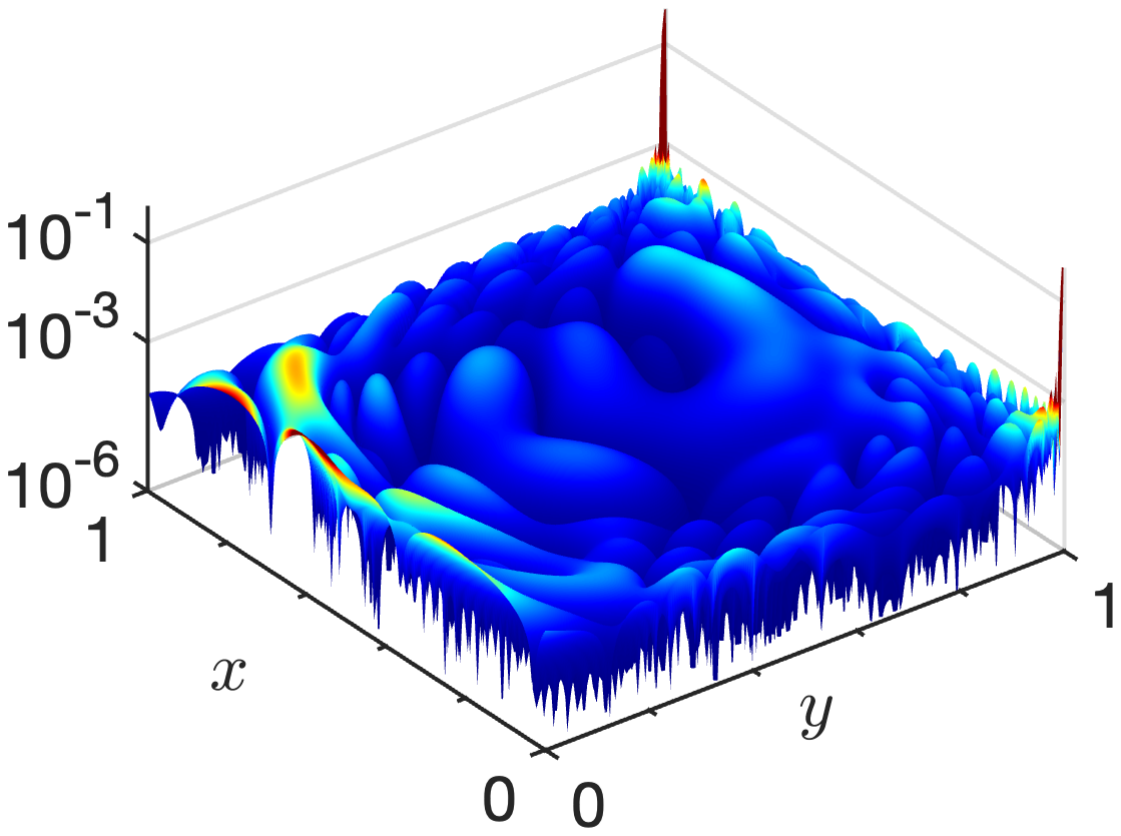}%
 \includegraphics[width=0.32\textwidth,trim=0 145 0 145,clip]{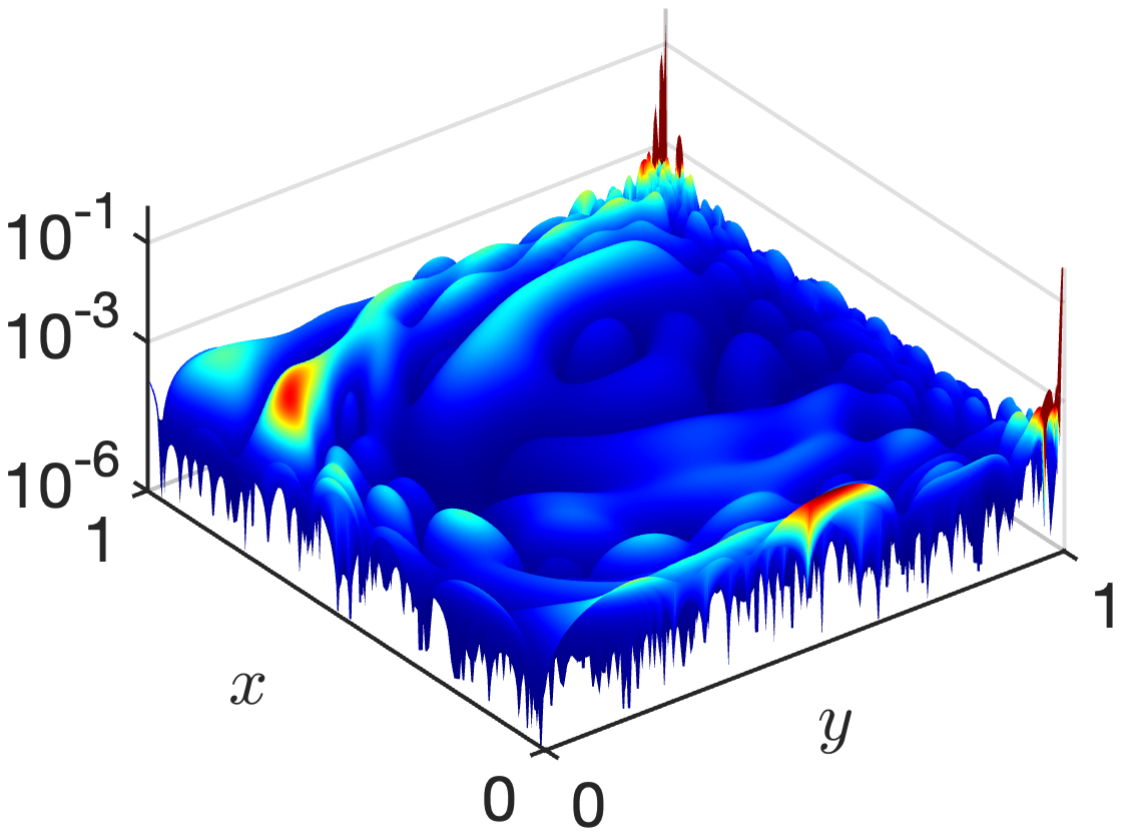}
  \includegraphics[width=0.32\textwidth,trim=0 145 0 145,clip]{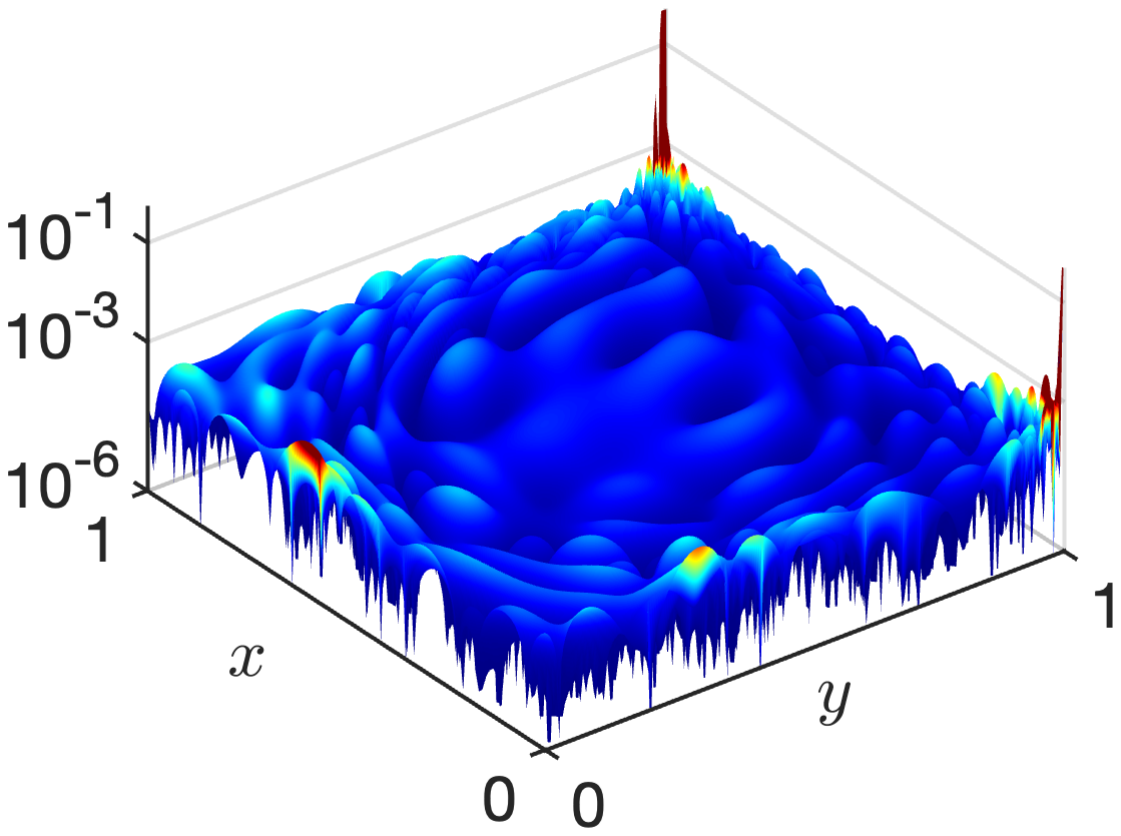}
  ~
  \subcaption{Case A, \\
  $\epsilon_u=89.0$, $\epsilon_v=88.6$, $\epsilon_p=94.0$}
\end{subfigure} 
~
~
\centering
  \begin{subfigure}{\textwidth}
\includegraphics[width=0.32\textwidth,trim=0 80 0 145,clip]{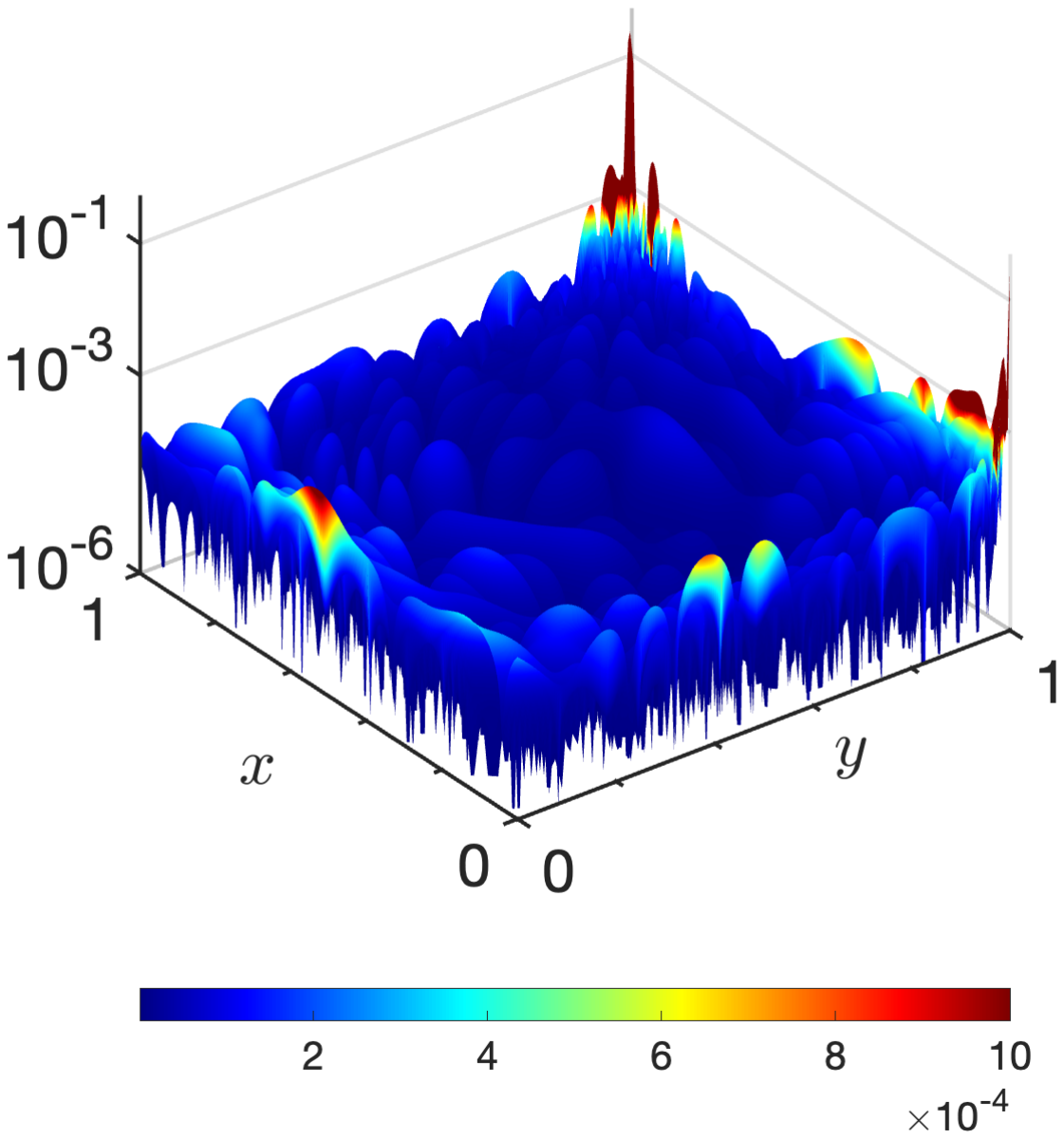}%
\includegraphics[width=0.32\textwidth,trim=0 80 0 145,clip]{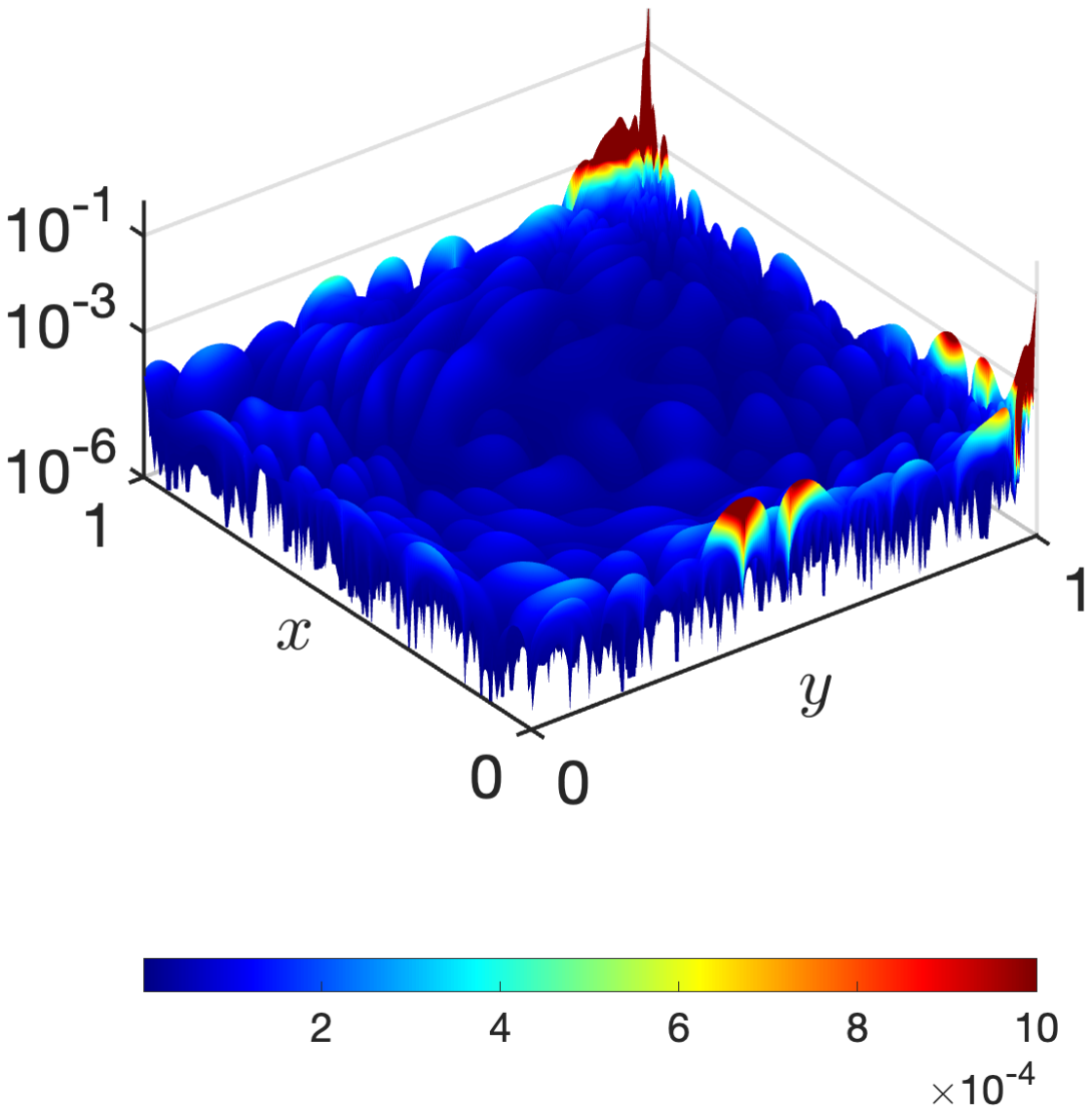}
\includegraphics[width=0.32\textwidth,trim=0 80 0 145,clip]{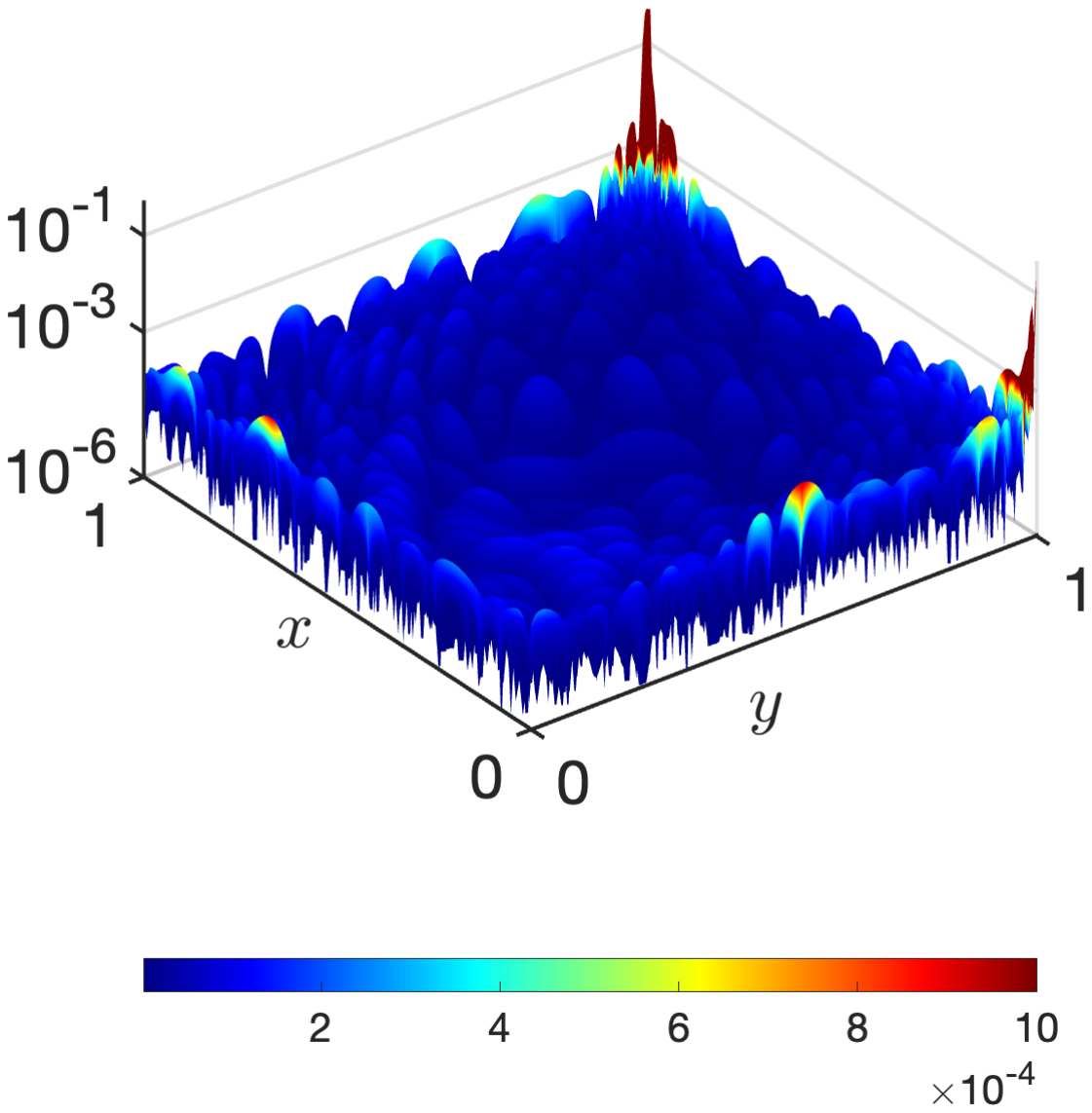}
~
\subcaption{Case B,\\$\epsilon_u=2.11$, $\epsilon_v=2.23$, $\epsilon_p=4.78$}
\end{subfigure} 
~
\caption{\textbf{Residues of the Navier-Stokes equations with NSFnet (upper) and ev-NSFnet (lower) reveal singularities at the corners.} Left panel: momentum residual in x direction; middle panel: momentum residual in y direction; right panel: velocity divergence. Note that in order to calculate the point-wise residues, we first infer $u, v, p$ from the trained NSFnet/ev-NSFnet and  then we use the second-order central finite difference formula to compute the derivatives on the lattices. Percentage errors are shown under each row. While the magnitudes of the residues are all comparable and close to zero, the errors with respect to the DNS solution are relatively large for the upper row, indicating the existence of another Navier-Stokes solution, different than the DNS solution. }
\label{fig:residual}
\end{figure}

In order to verify that the results we obtained are valid solutions, we substituted the inferred $u,\, v$ and $p$ of case A1 and case B5 into the 2D steady Navier-Stokes equation \ref{NSE1} and \ref{NSE2}, and the resulting point-wise residues of each equation are plotted in Figure \ref{fig:residual}. We observe that the magnitudes of the residues of all equations are less than $10^{-3}$, except at the two corners at $x=0,\,y=1$ and $x=1,\, y=1$. Although the solution of \emph{class 1} satisfies both the governing equations and boundary conditions, the RPE of case A1 is greater than 90\%, therefore it could be inferred that this is another valid solution for the 2D steady cavity flow at $Re=2,000$. 
\begin{figure*}
\captionsetup[subfigure]{justification=centering}
\centering
\begin{subfigure}{0.43\textwidth}
\includegraphics[width=\textwidth,trim=0 125 0 145,clip]{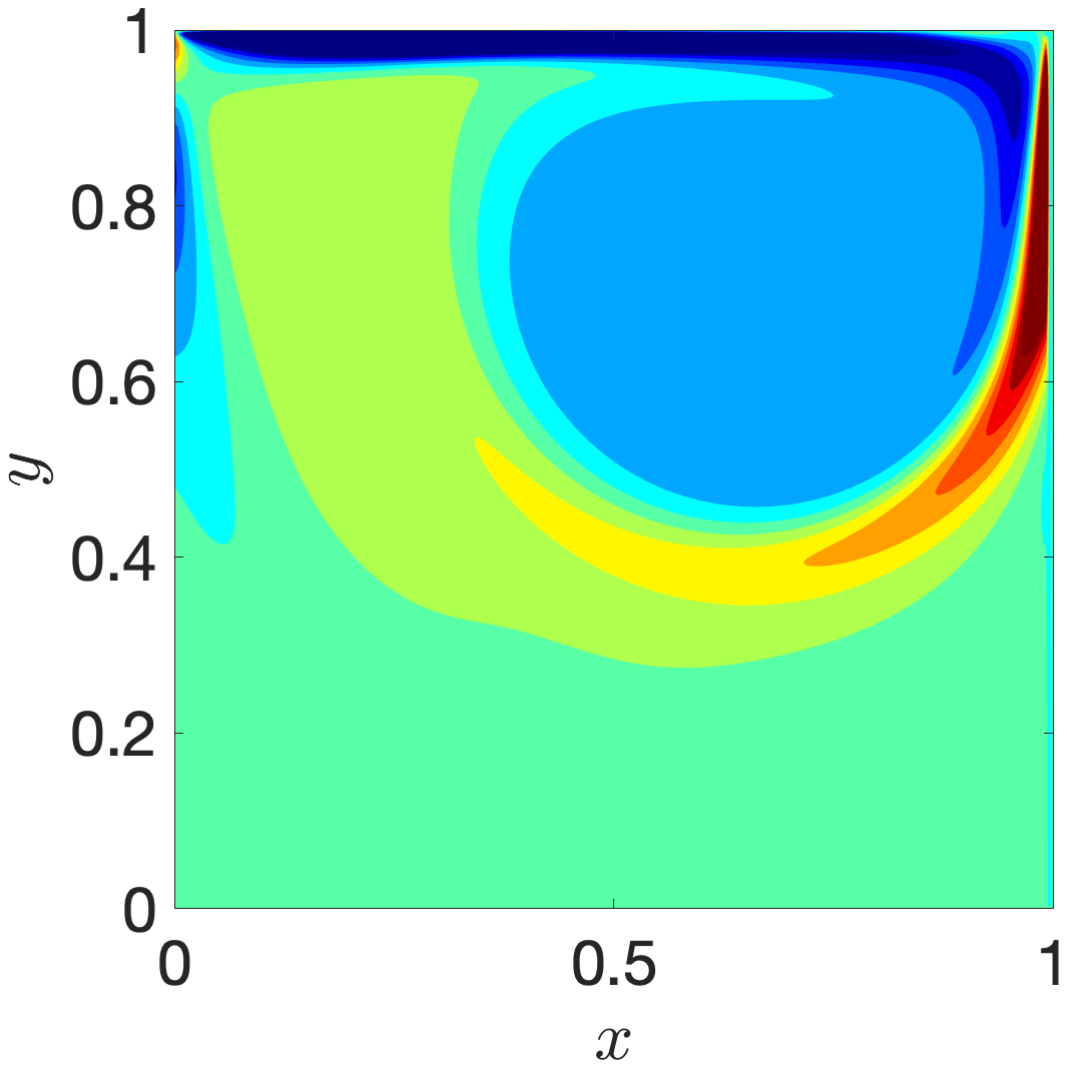}  
  \subcaption{Case A, \\
  $\scriptstyle \epsilon_u=89.0$, $\scriptstyle\epsilon_v=88.6$, $\scriptstyle \epsilon_p=94.0$.
   }
\end{subfigure} 
~
\begin{subfigure}{0.475\textwidth}
\includegraphics[width=\textwidth,trim=0 145 0 145,clip]{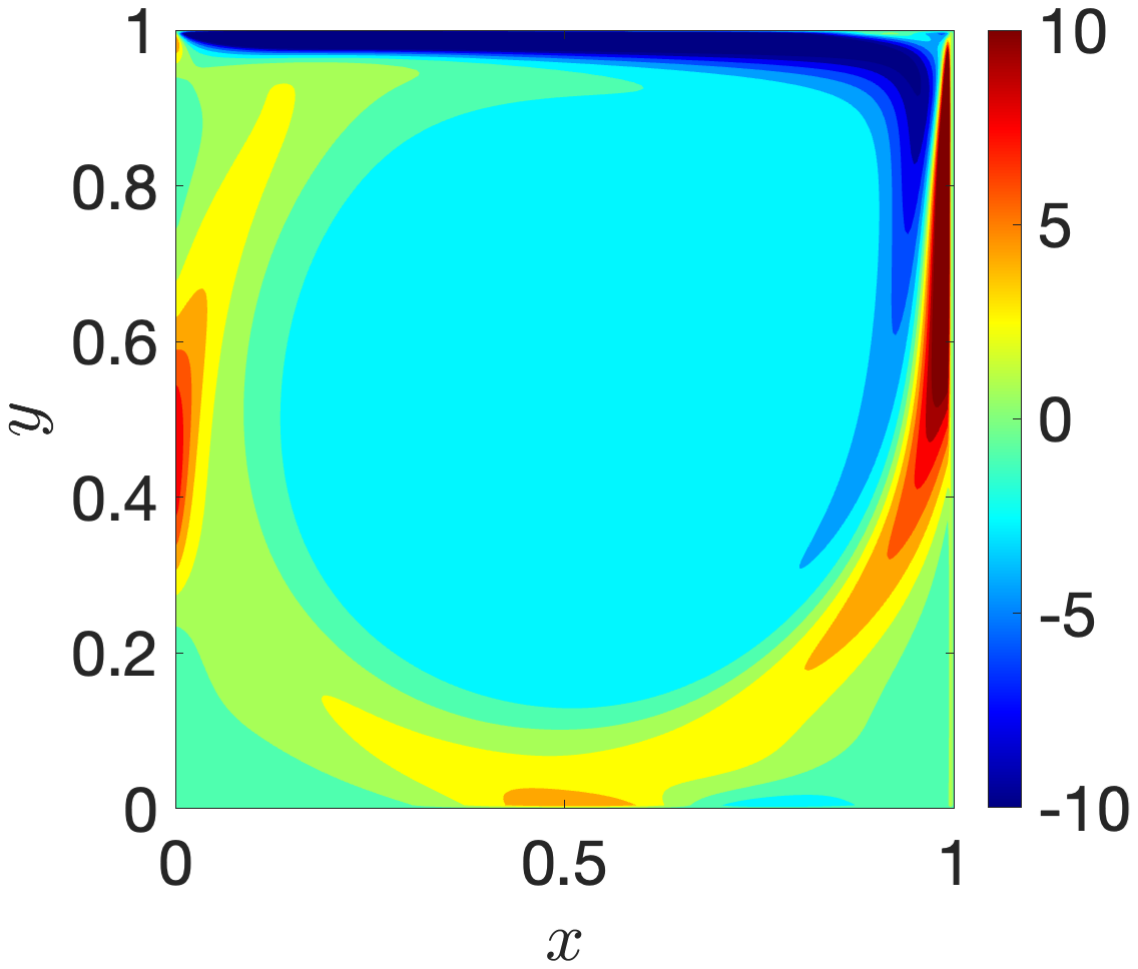}  
  \subcaption{Case B
  ,\\$ \scriptstyle \epsilon_u=2.11$, $\scriptstyle \epsilon_v=2.23$, $\scriptstyle \epsilon_p=4.78$
  }
\end{subfigure} 
~
\caption{\textbf{Distribution of the vorticity of the steady cavity flow at $\boldsymbol{Re = 2,000}$.}  NSFnet (left) and ev-NSFnet (right). The result on the right plot is obtained by the parameterized model using a single network structure. The large difference in the vorticity pattern indicates that these are two different Navier-Stokes solutions. Corresponding percentage errors are shown under each plot.}
\label{fig:vorticity}
\end{figure*}

Furthermore, in order to verify that the solution of \emph{class 1} is different from the DNS solution, we plot the distribution of the vorticity derived from the inference in Figure \ref{fig:vorticity}, which clearly depicts that the solution of \emph{class 1} (left figure) is totally different from the reference one (right figure), obtained by DNS.

\begin{figure*}
\captionsetup[subfigure]{justification=centering}
\centering
\begin{subfigure}{0.43\textwidth}
\includegraphics[width=\textwidth,trim=0 125 0 145,clip]{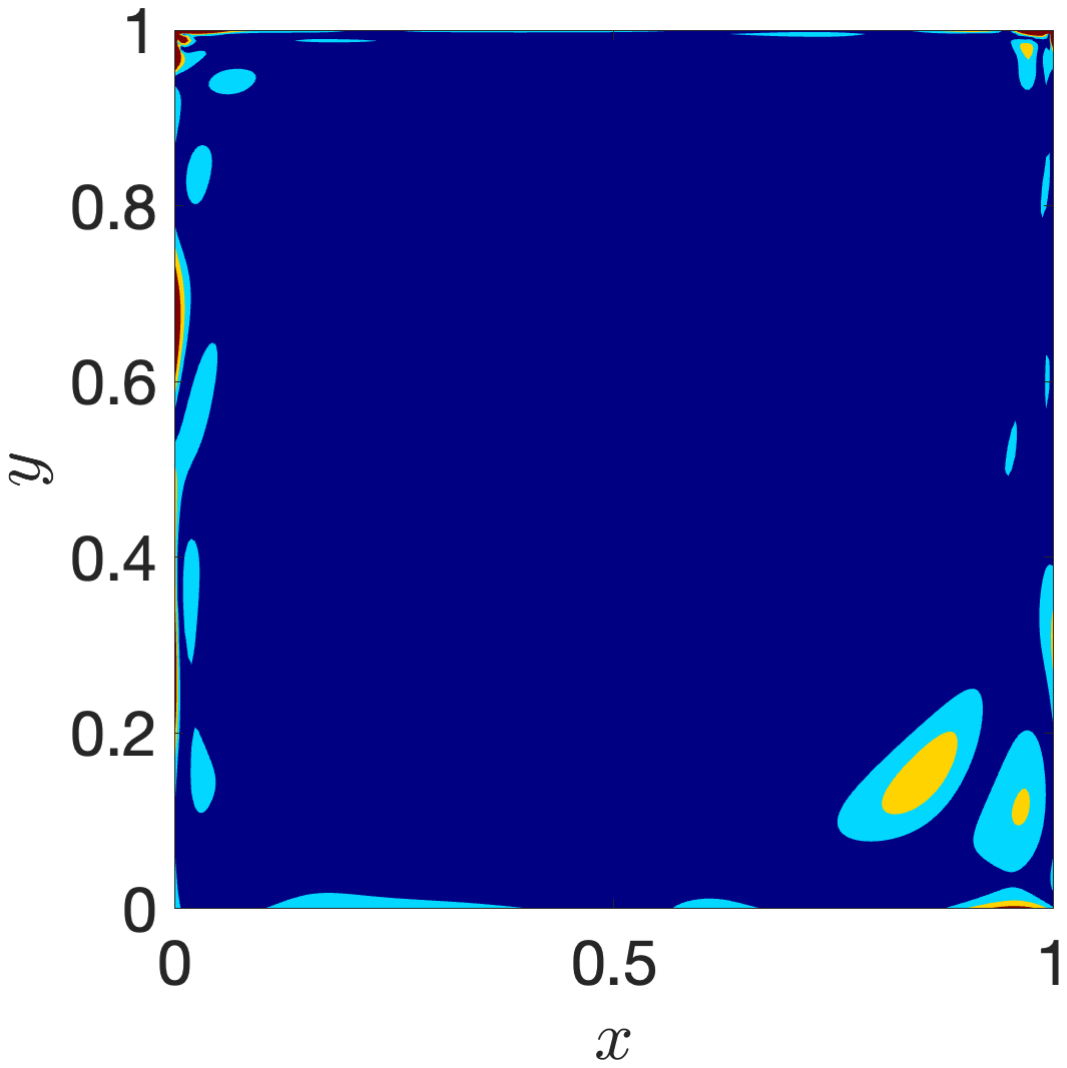}  
  \subcaption{
  Case B, \\
  $Re=2\,000$}
\end{subfigure} 
~
\begin{subfigure}{0.475\textwidth}
\includegraphics[width=\textwidth,trim=0 145 0 145,clip]{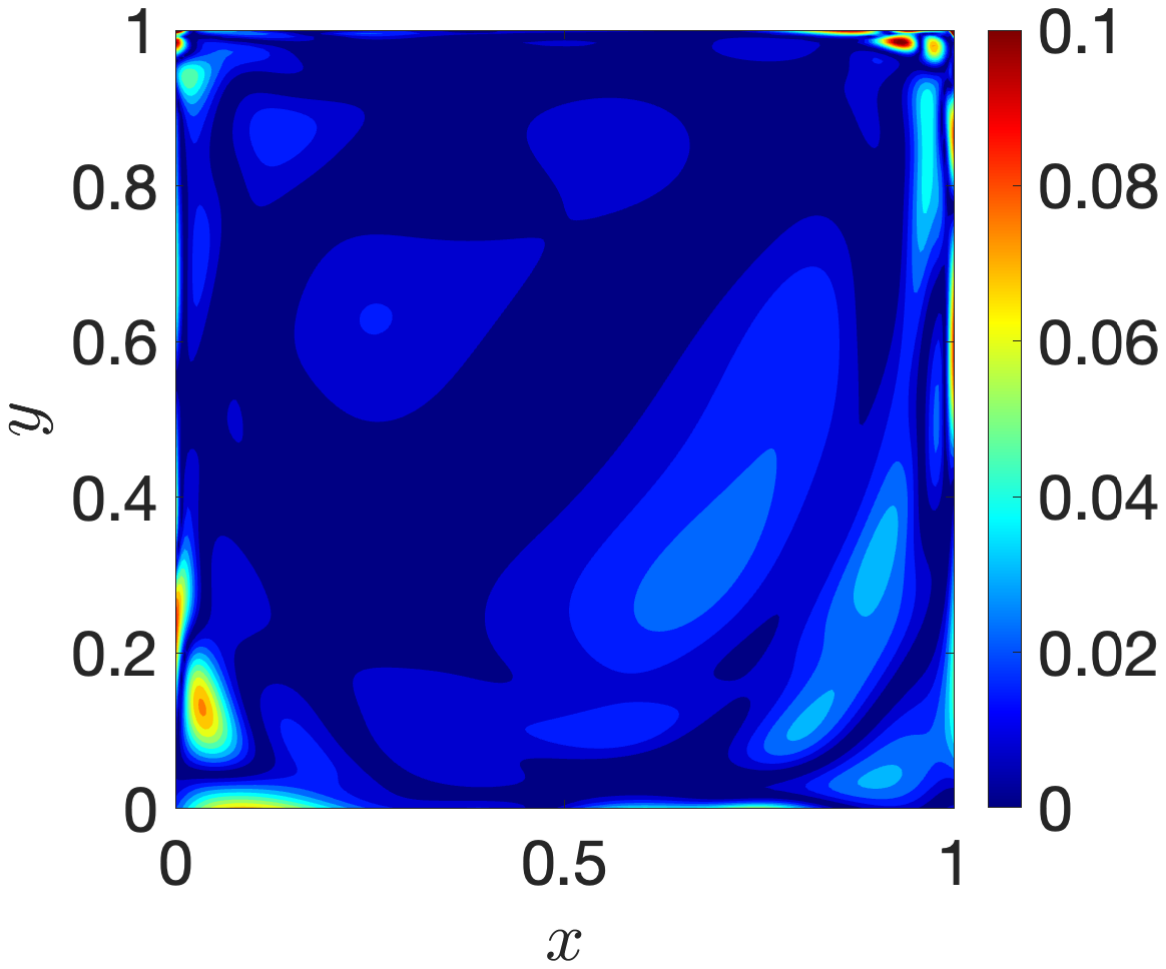}  
  \subcaption{
  Case C, \\
  $Re=3\,000$}
\end{subfigure} 
~
\caption{\textbf{Distribution of the normalized eddy viscosity (divided by the molecular viscosity)  of the steady cavity flow at $\boldsymbol{Re = 2,000}$ and $\boldsymbol{Re = 3 000}$.} Note that in the training, the parameterized model is used without labeled data.  Larger viscosity values are observed at the corners and the walls. In the left plot, we use 40,000 residual points, while in the right plot we use 60,000 residual points. The network parameters as well as the eddy viscosity model parameters are the same for both cases.
}
\label{fig:evm_viscosity}
\end{figure*}

\section{Effects of eddy viscosity and data}
The aforementioned results show that the entropy viscosity can help the PINNs optimizer to select the `correct' class of solutions and improve the inference accuracy substantially for the steady cavity flow at high $Re$. However, as shown by equations \ref{eq:loss1} and \ref{eq:loss2}, if the magnitude of $\nu_E$ has a big value compared to the molecular viscosity $\nu$, the optimization problem degenerates to cavity flow at a lower $Re$. Fortunately, as demonstrated by Figure \ref{fig:evm_viscosity}, the maximum value of eddy viscosity given by ev-NSFnet is less than $\frac{1}{10}\nu$ in the majority of the domain, except in the region close to the walls, in particular in the corners where singularity appear. 

\begin{figure}
\centering
~
\begin{subfigure}[ht!]{0.9\textwidth}
\includegraphics[width=0.475\textwidth]{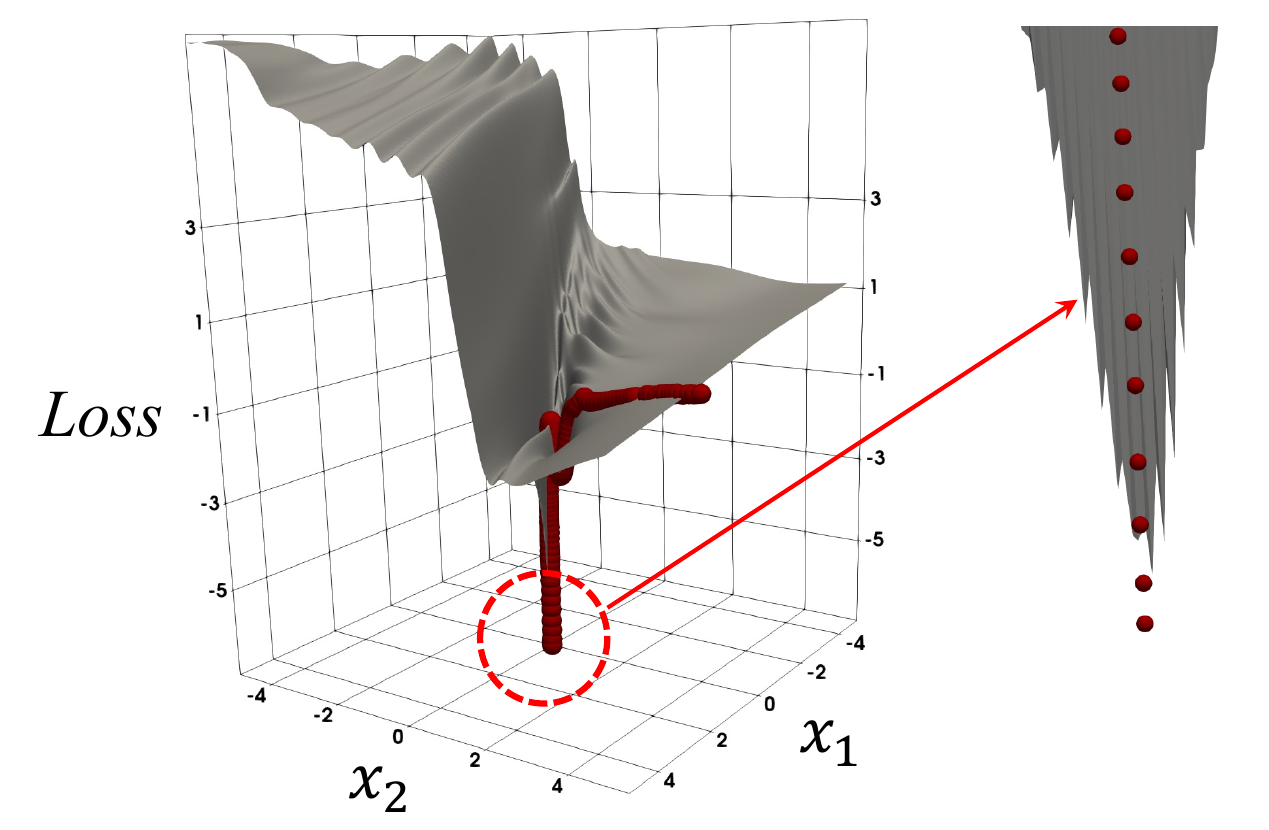}%
\includegraphics[width=0.475\textwidth,trim=50 190 50 210,clip]{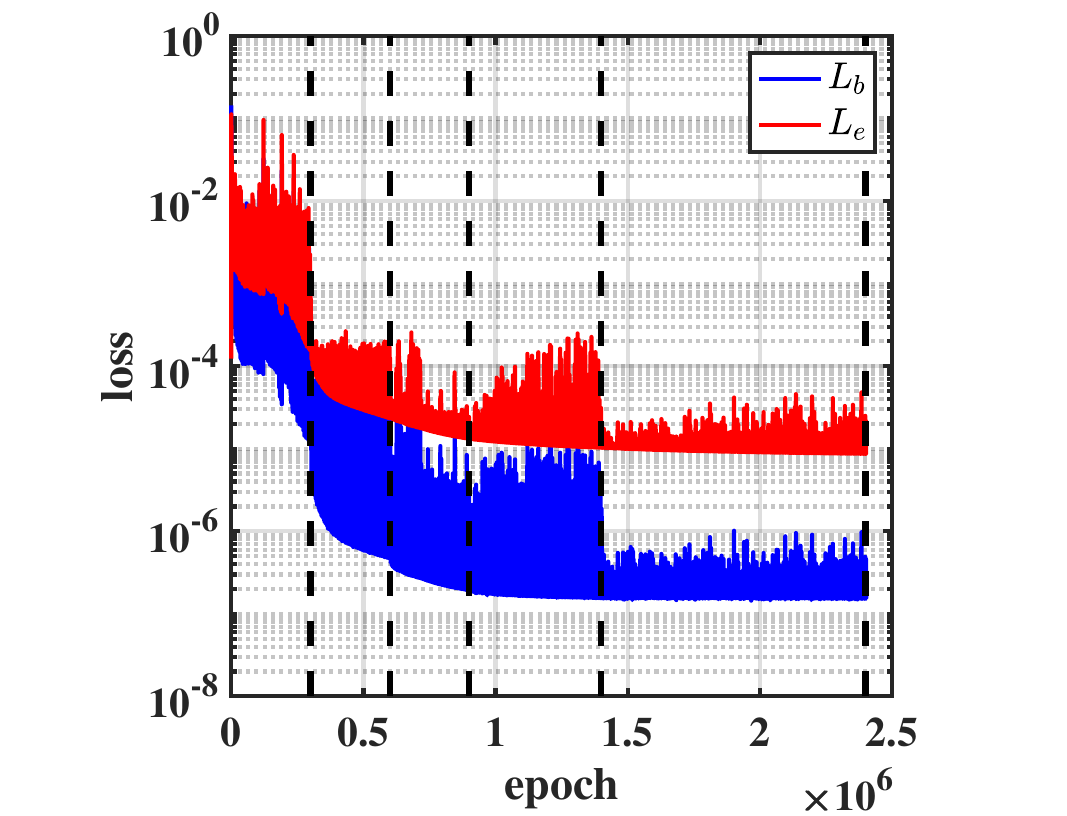}%
 \subcaption{Case A, \\
  $\epsilon_u=89.0$, $\epsilon_v=88.6$, $\epsilon_p=94.0$}
\end{subfigure} 
~
 \begin{subfigure}{0.9\textwidth}   \includegraphics[width=0.475\textwidth,height=0.325\textwidth, trim=0 0 0 0,clip]{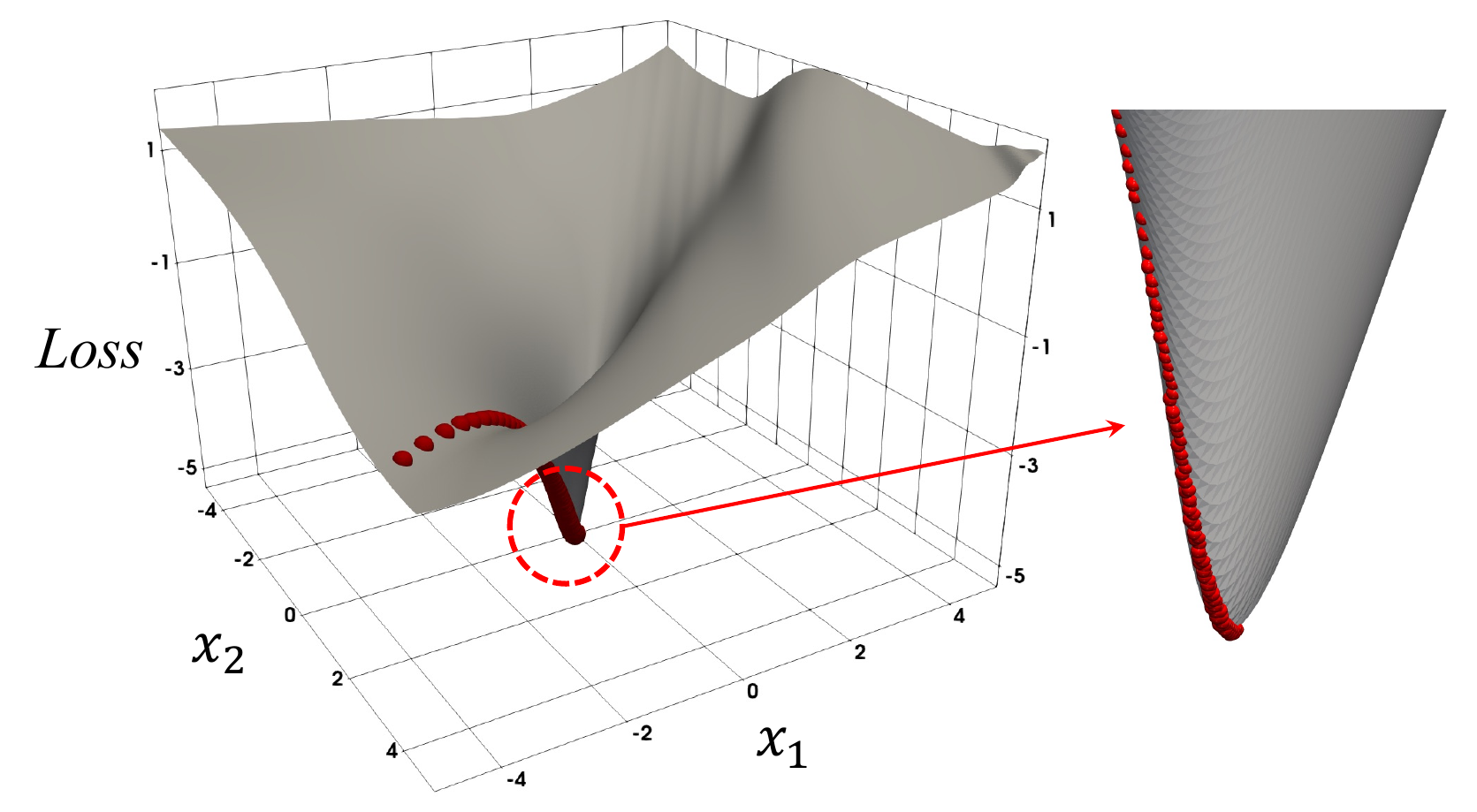}
   \includegraphics[width=0.475\textwidth,trim=50 190 50 210,clip]{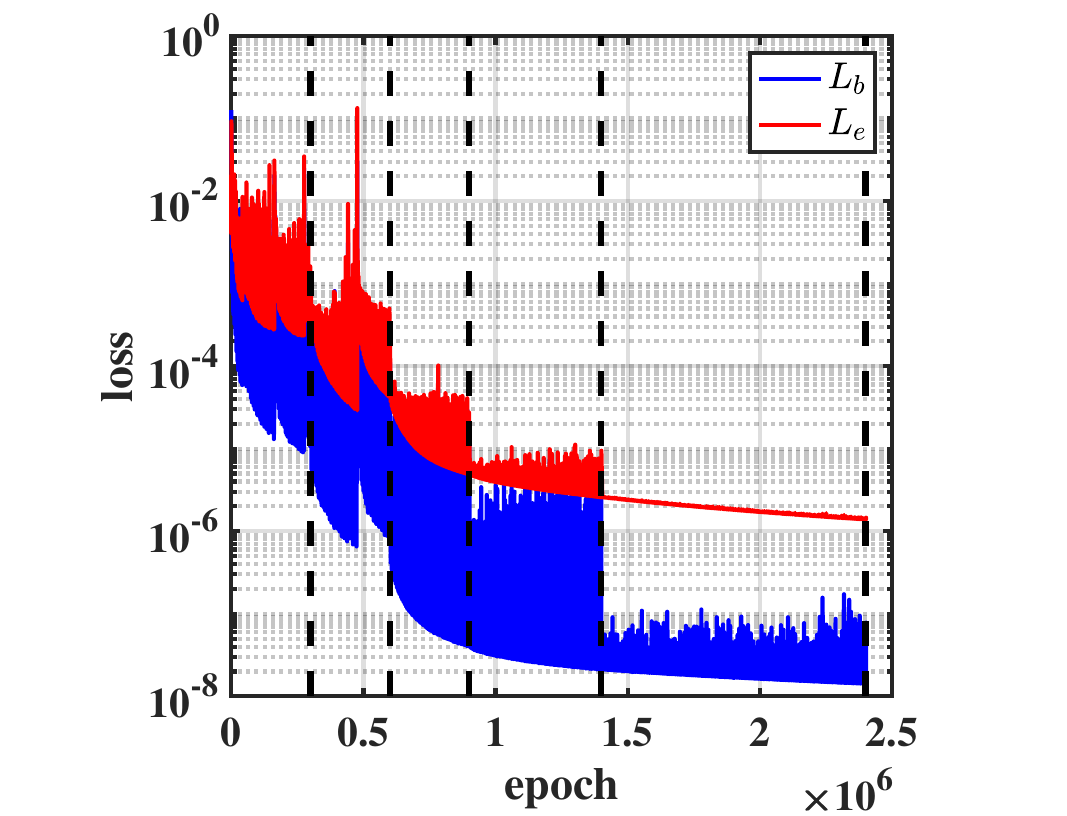}%
   \subcaption{ Case B, \\
  $\epsilon_u=0.89$, $\epsilon_v=0.92$, $\epsilon_p=1.42$}
  \end{subfigure} 
~
\centering
 \begin{subfigure}{0.9\textwidth}   \includegraphics[width=0.475\textwidth,height=0.325\textwidth, trim=0 0 0 0,clip]{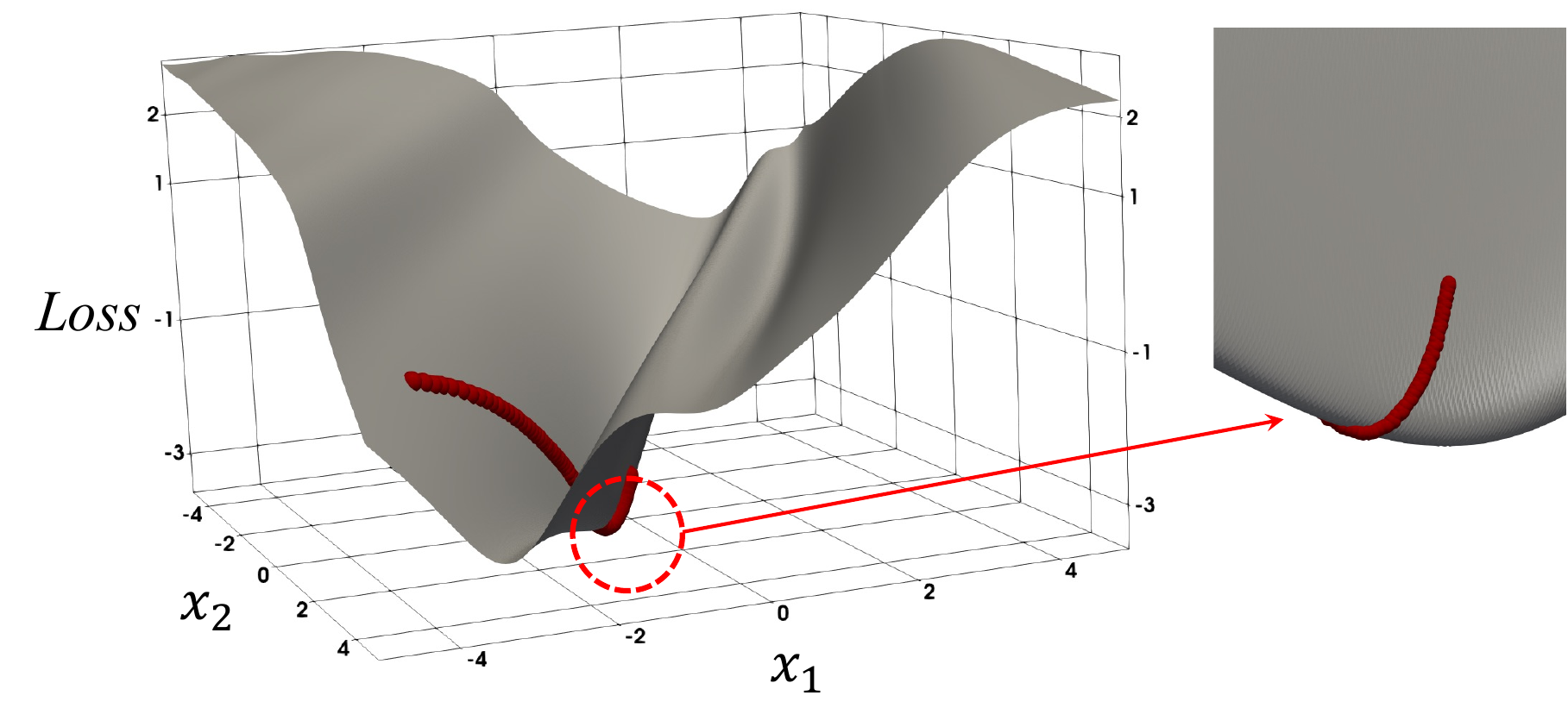}
   \includegraphics[width=0.475\textwidth,trim=50 190 50 210,clip]{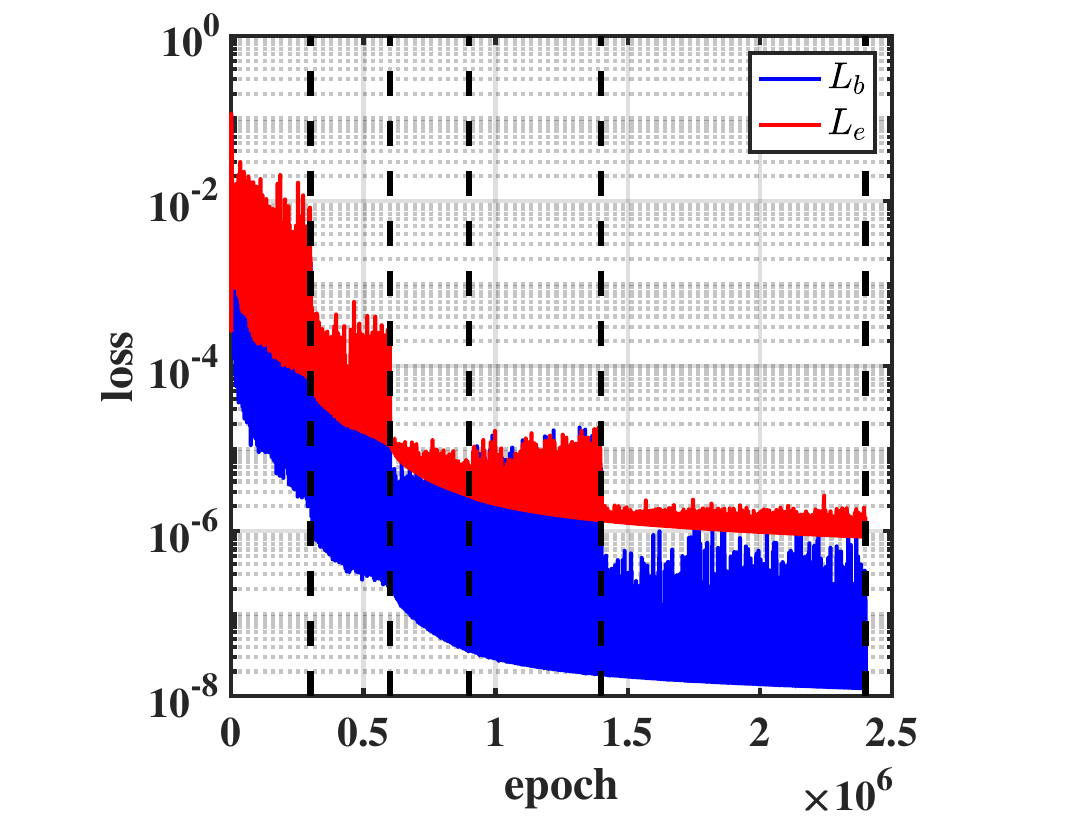}%
\subcaption{Case C,\\
  $\epsilon_u=2.1$, $\epsilon_v=2.2$, $\epsilon_p=4.4$}
\end{subfigure} 
~
\caption{\textbf{Loss landscape and loss history by the NSFnet/ev-NSFnet for $\boldsymbol{Re = 2,000}$.} (top) unstable solution, (middle) one labeled data is used in training; (bottom) no labeled data but the parameterized
eddy-viscosity model is used. Both the labeled data and the eddy viscosity smooth the loss landscape, and the optimizer approaches the global minimum corresponding to the DNS stable solution easier. Note that with only one data point the error in the velocity field is less than 1\%. In Case C (middle plot),  one labeled data at ($x=0.7,\,y=0.5$) is used.}
\label{fig:landscape_Re2K}
\end{figure}
Further understanding of the effect of eddy viscosity and labeled data on model training can be achieved by visualising the loss landscapes and training trajectories. Following the work by \citep{Li2017VisualizingTL,NEURIPS2021_df438e52}, the objective loss function $L$ can be reformulated as follows,
\begin{align}      f(x_1,x_2)=L(W_n+x_1 \xi+x_2 \gamma),
\end{align}
where $\xi$ and $\gamma$ are the direction vectors corresponding to the first two components by Principal Component Analysis (PCA) of matrix $[W_1-W_n, W_2-W_n, ..., W_{n-1}-W_n]$, where subscript $n$ denotes the $n^{th}$ epoch; $x_1$ and $x_2$ are the directional weights associated to $\xi$ and $\gamma$, respectively.
 
Using landscape plots we can further demonstrate that the accuracy improvement can be attributed to eddy viscosity and labeled data. Figure \ref{fig:landscape_Re2K} exhibits the loss landscape of three cases, namely, case A: no labeled data; case C: one labeled data at $x=0.7,y=0.5$ but no eddy viscosity; case B: no labeled data but parameterized eddy viscosity model. As shown in Figure \ref{fig:landscape_Re2K} (a), without regularization by the labeled data or the eddy viscosity, the loss surface is bumpy, and notable roughness can be observed as we zoom into the loss surface, while the MSE loss can barely decrease to $10^{-5}$. When one labeled data (e.g., single point measurement) is used in the training, as shown in Figure \ref{fig:landscape_Re2K} (b), the surface roughness of the loss lanscae disappears and the MSE loss of the equations is reduced to $10^{-6}$. Furthermore, when no labeled data is available but eddy viscosity is introduced instead, as shown in Figure \ref{fig:landscape_Re2K} (c), the loss surface becomes smooth and the MSE loss of the equations is reduced to less than $10^{-6}$.
\begin{figure}
~
\centering
\begin{subfigure}{0.99\textwidth}
\includegraphics[width=0.35\textwidth,trim=0 145 0 145,clip]{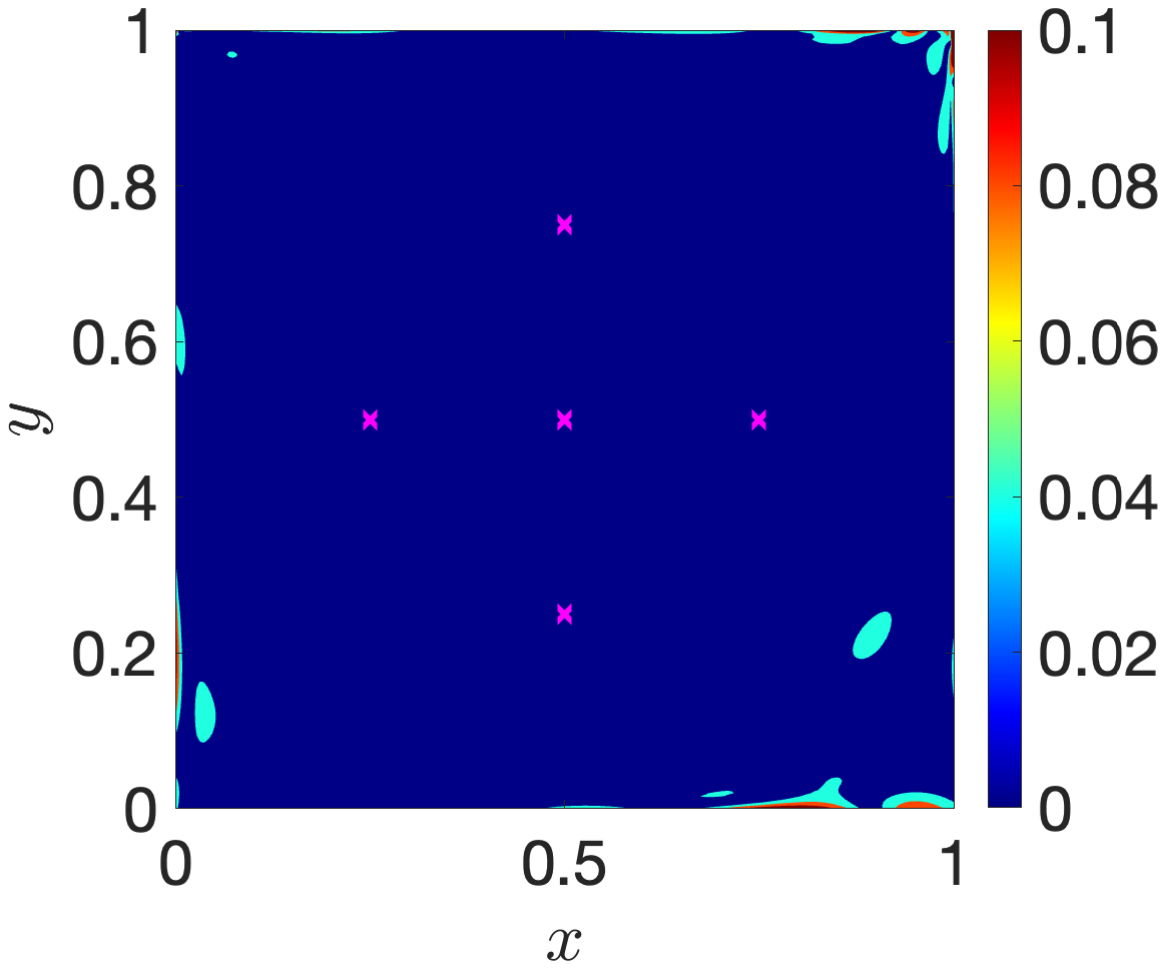}  \includegraphics[width=0.32\textwidth,height=0.22\textwidth,trim=0 -50 0 0,clip]{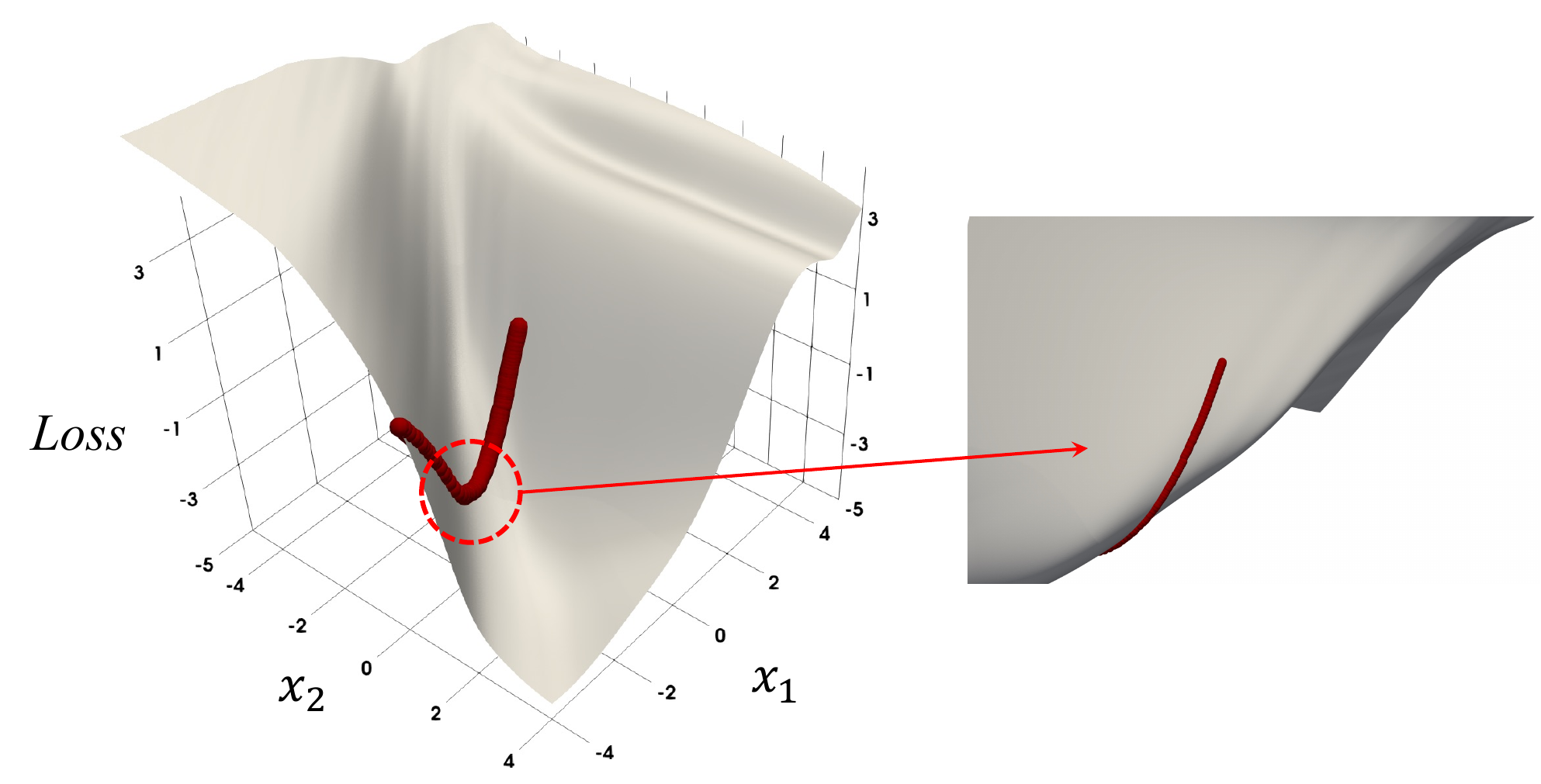}
\includegraphics[width=0.32\textwidth,trim=50 190 50 210,clip]{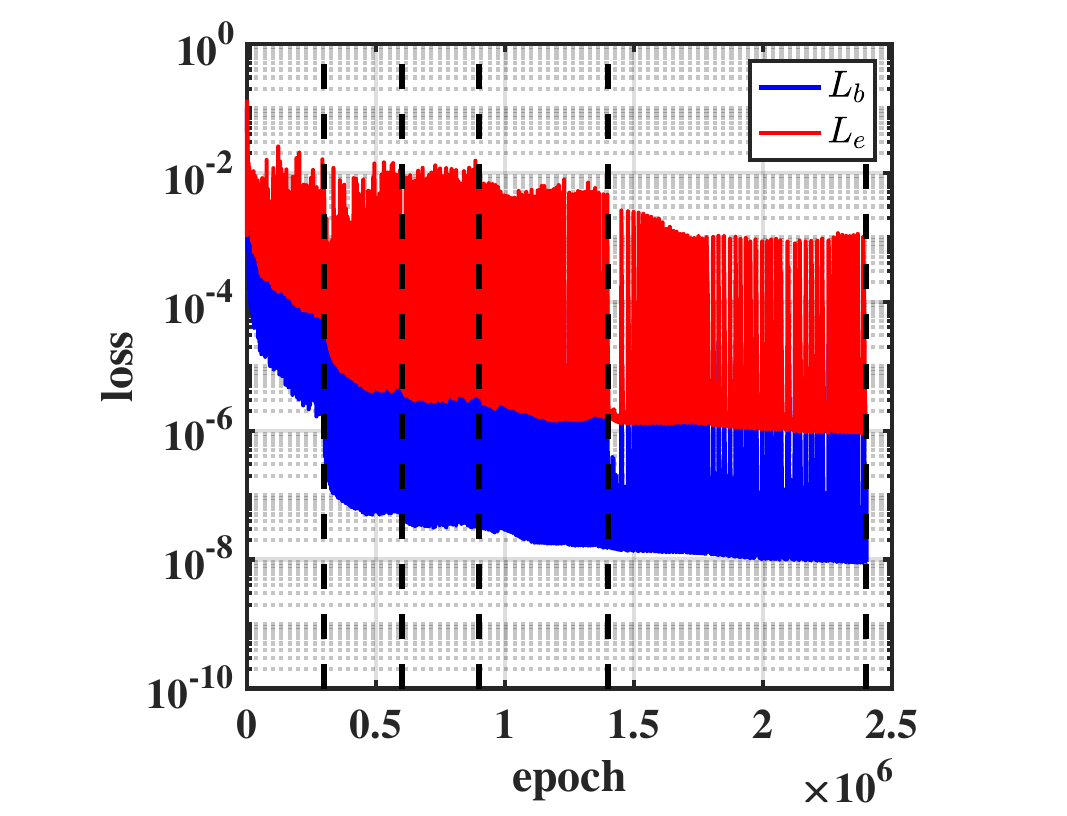}%
\subcaption{Case E, \\
\emph{single-network parameterized}; 5 data point, \\
  $\epsilon_u=0.54$, $\epsilon_v=0.51$, $\epsilon_p=2.55$.}
\end{subfigure} 
~
\centering
\begin{subfigure}{0.99\textwidth}
\includegraphics[width=0.35\textwidth,trim=0 145 0 145,clip]{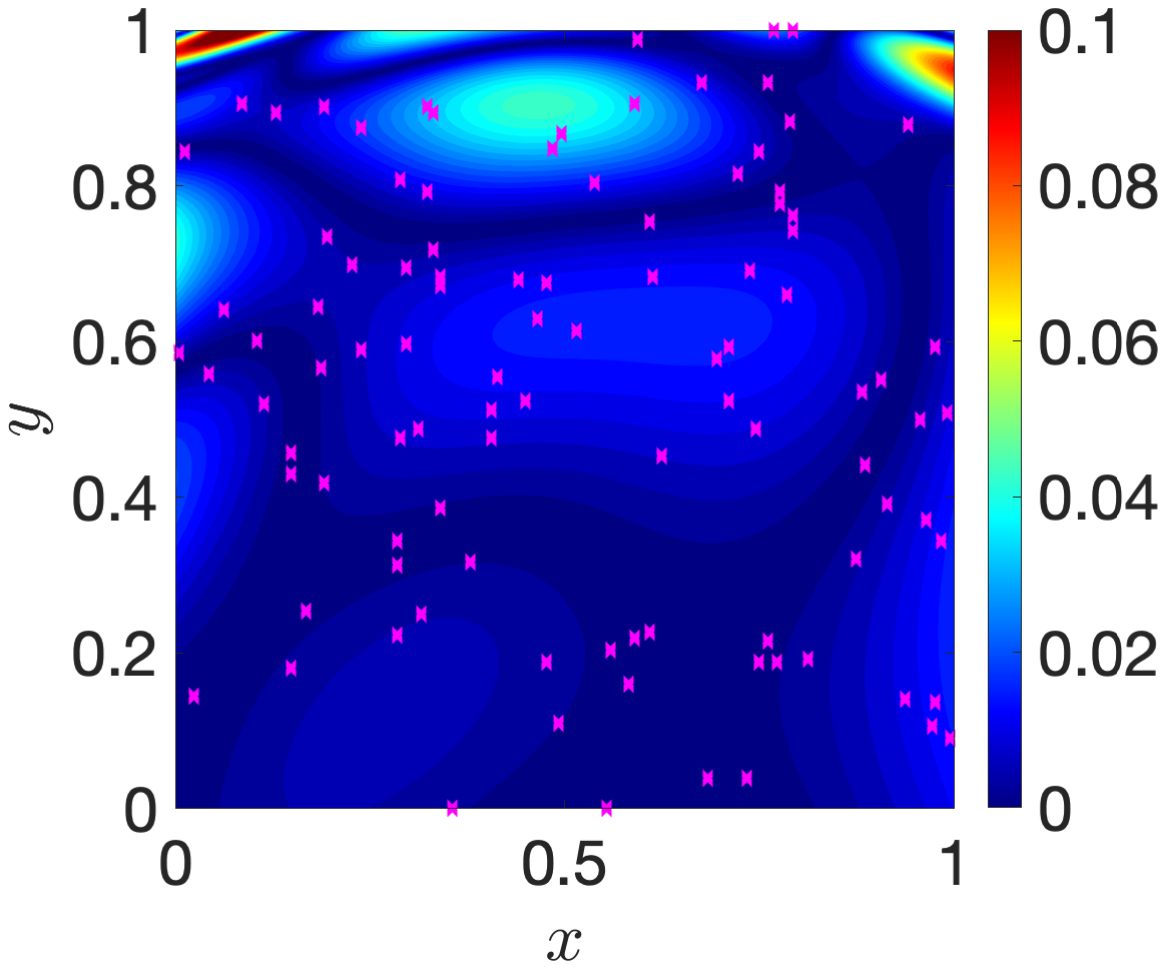}  \includegraphics[width=0.32\textwidth,height=0.22\textwidth,trim=0 -50 0 0,clip]{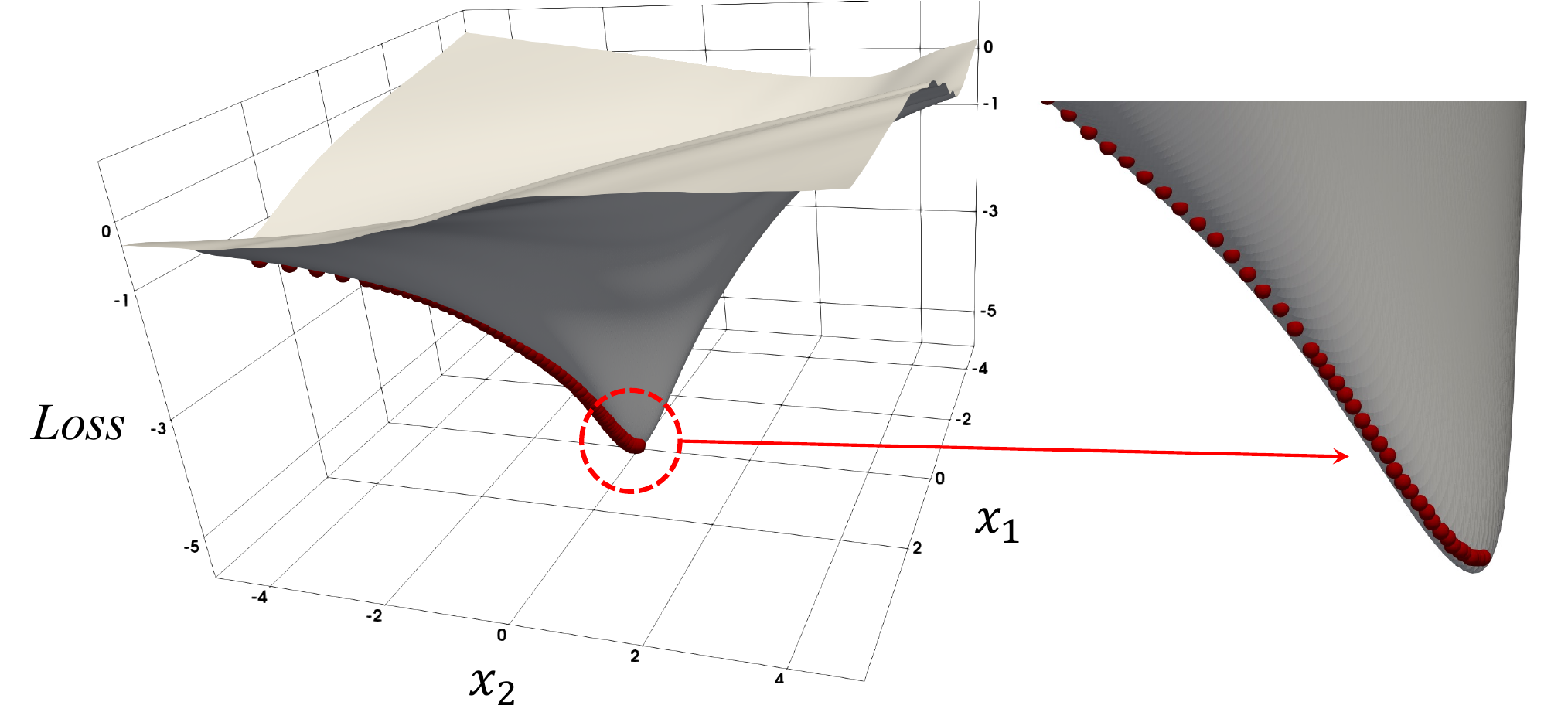}   \includegraphics[width=0.32\textwidth,trim=50 190 50 210,clip]{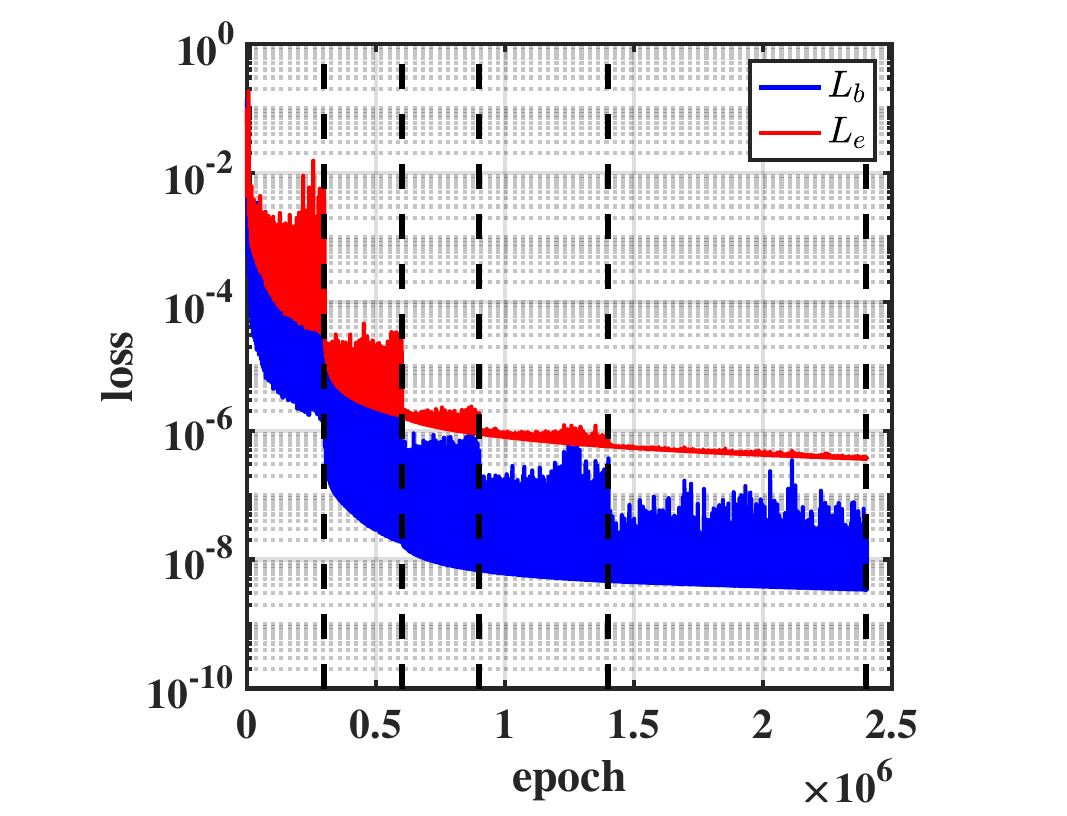}%
  \subcaption{Case F; \\
  \emph{neural network model}; 100 data point; \\
  $\epsilon_u=0.27$, $\epsilon_v=0.26$, $\epsilon_p=0.73$.}
\end{subfigure} 
~
\caption{\textbf{Learning eddy viscosity model directly from data.} Comparison of the normalized eddy viscosity obtained by the optimally parameterized model and the neural network model, when a number of labeled data within the computational domain are available (pink points), Re = 2 000. The neural network on the right produces an optimum eddy viscosity distribution and thus the velocity error with respect to DNS is smaller. The loss landscapes in both cases are smooth, but the lower case with neural network model for eddy viscosity leads to smaller velocity errors. The pink cross denotes the location of labeled data. 
}
\label{fig:eddy_viscosity_inv}
\end{figure}

Therefore,  we can conclude that the equation residual based eddy viscosity can smooth the loss landscape and guide the PINNs optimizer to seek the global minimum, which gives a solution close to DNS. However, in the scenario that a few labeled data are available, whether the eddy viscosity can be learned from the data is not clear. To this end, we have also considered two new cases that utilize both the eddy viscosity and the labeled data. In particular, we introduce a separate  \emph{neural network model} for the eddy viscosity in the case where 100 randomly selected labeled data points are available; in this case we learn the eddy viscosity as a function of $(x,y)$, without using the parametric eddy viscosity model. Figure \ref{fig:eddy_viscosity_inv} (a) presents the learned eddy viscosity through the \emph{parameterized model}, where 5 points are used to estimate the eddy viscosity parameters. We observe that the learned eddy viscosity is mostly concentrated on the top two corners of singularity, similar to distribution in the case of no labeled data. The loss landscape is smooth as well, but the MSE loss history shows a significant number of spikes, which can be attributed to over-fitting of the \emph{parameterized model} as shown in Equation \ref{eq:nu_e}, where only two free parameters need to be determined. On the other hand, when the \emph{neural network model} is used, similar to that of the \emph{parameterized model}, the learned eddy viscosity is mostly distributed in the top two corners of singularity and the loss landscape is smooth, but the MSE loss history does not show many spikes, which indicates that the \emph{neural network model} is more flexible when a few labeled data are known. Note that with a few of labeled data both cases improve the inference accuracy for the velocity field to RPE less than 1\%.
\begin{figure}[ht!]
\centering
  \begin{subfigure}{0.95\textwidth}  \includegraphics[width=0.32\textwidth,trim=0 125 0 145,clip]
  {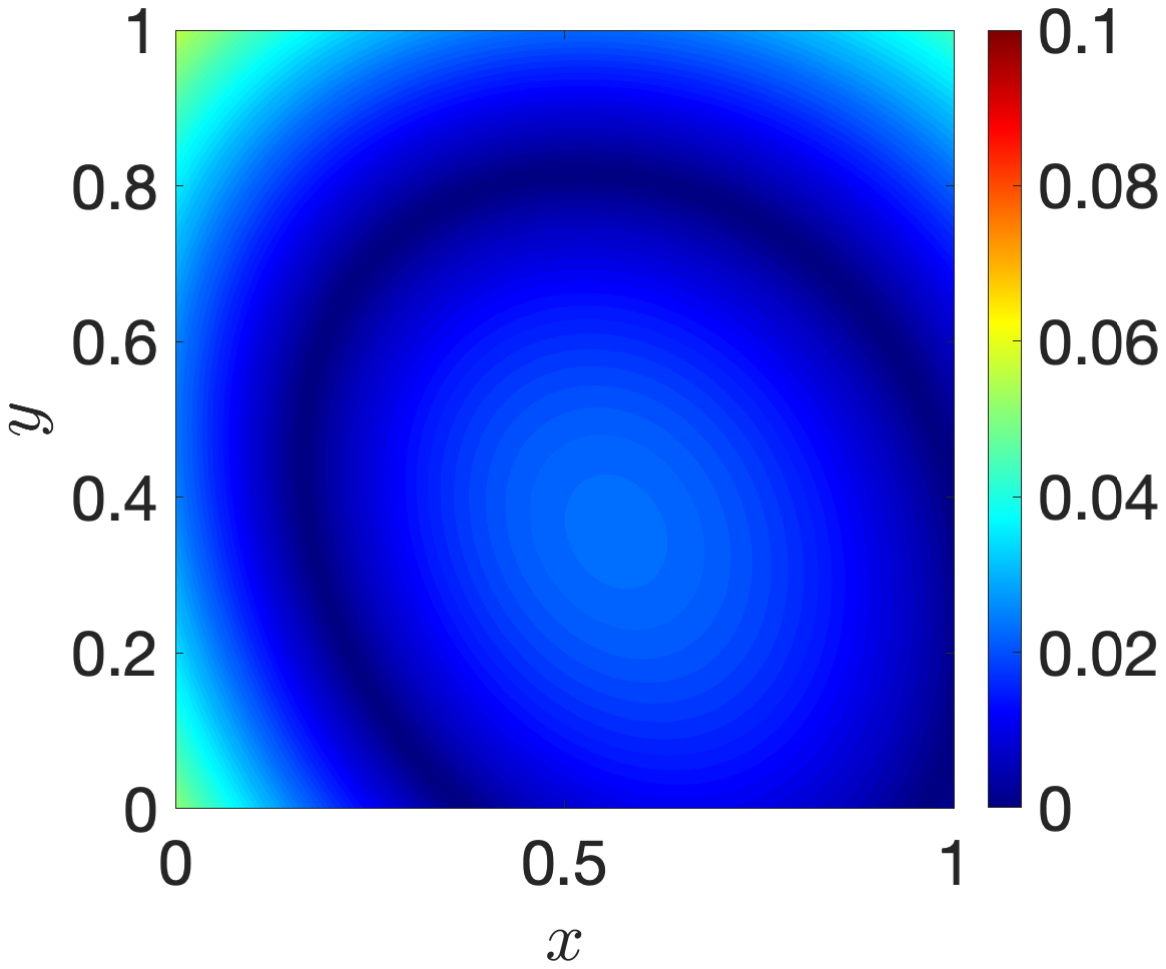}  
\includegraphics[width=0.32\textwidth,trim=1 125 1 125,clip]{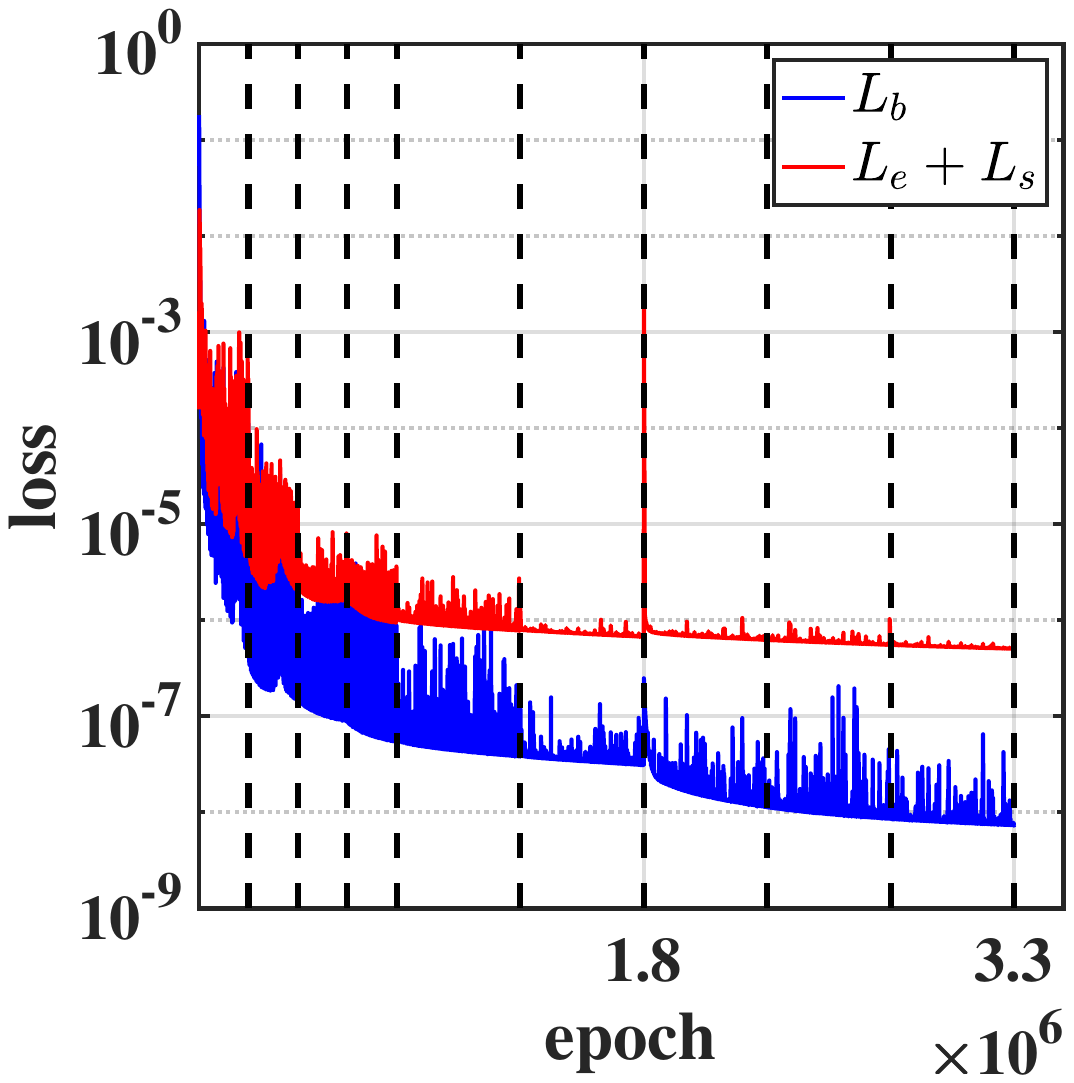}
\includegraphics[width=0.32\textwidth,trim=1 125 1 125,clip]{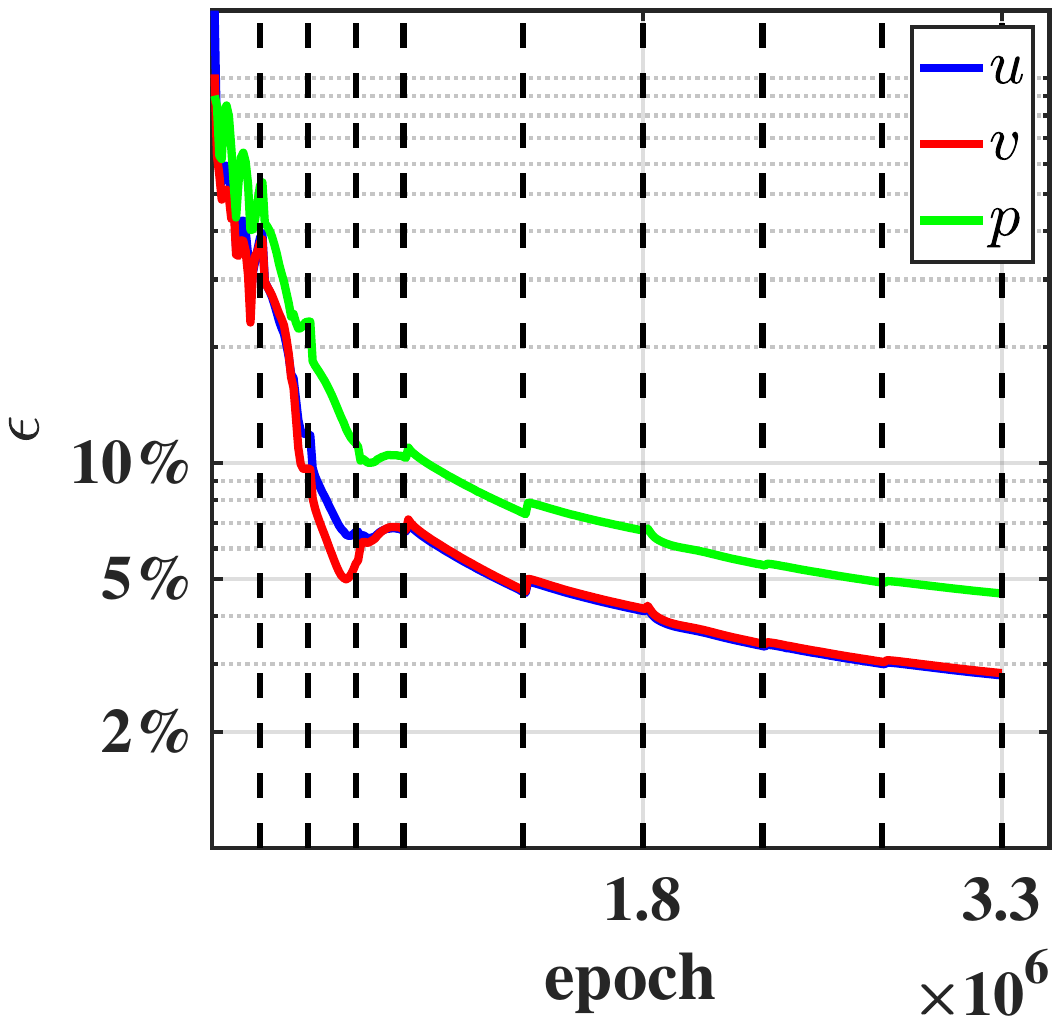}
\subcaption{Case G: \\ $Re=3\,000$}
\centering
\end{subfigure} 
~
 \begin{subfigure}{0.95\textwidth}   \includegraphics[width=0.32\textwidth,trim=0 145 0 145,clip]{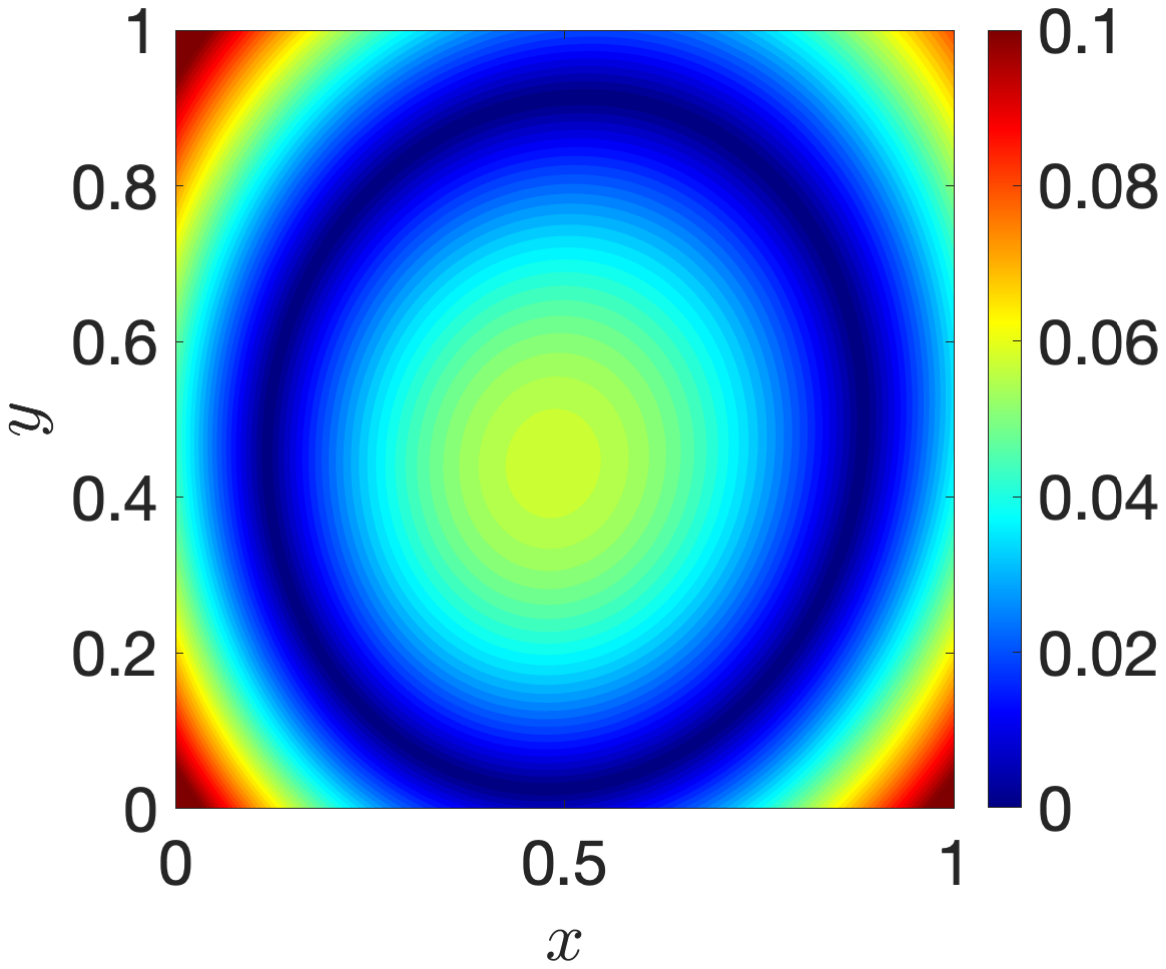} \includegraphics[width=0.32\textwidth,trim=1 125 1 125,clip]{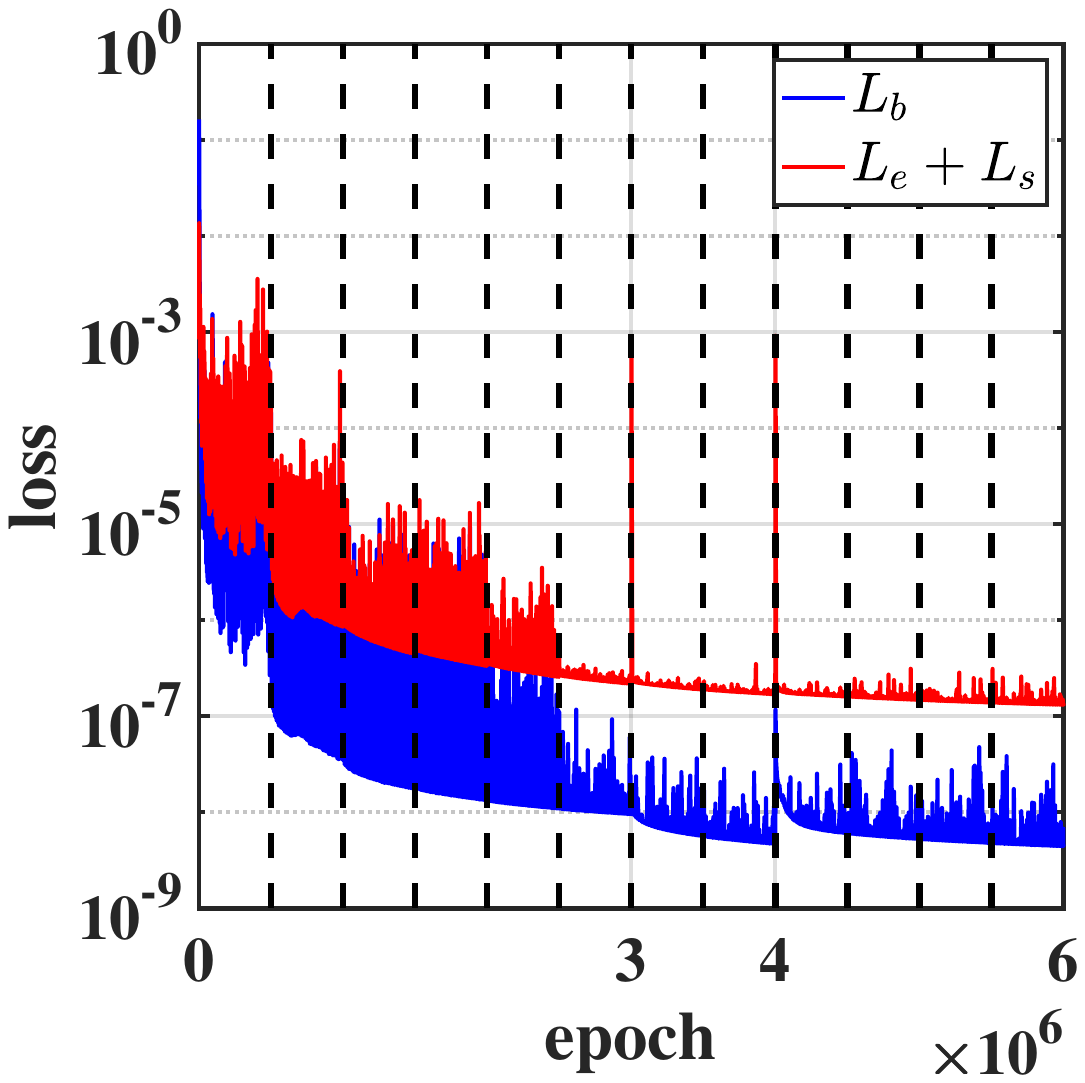}
   ~   \includegraphics[width=0.32\textwidth,trim=1 125 1 125,clip]{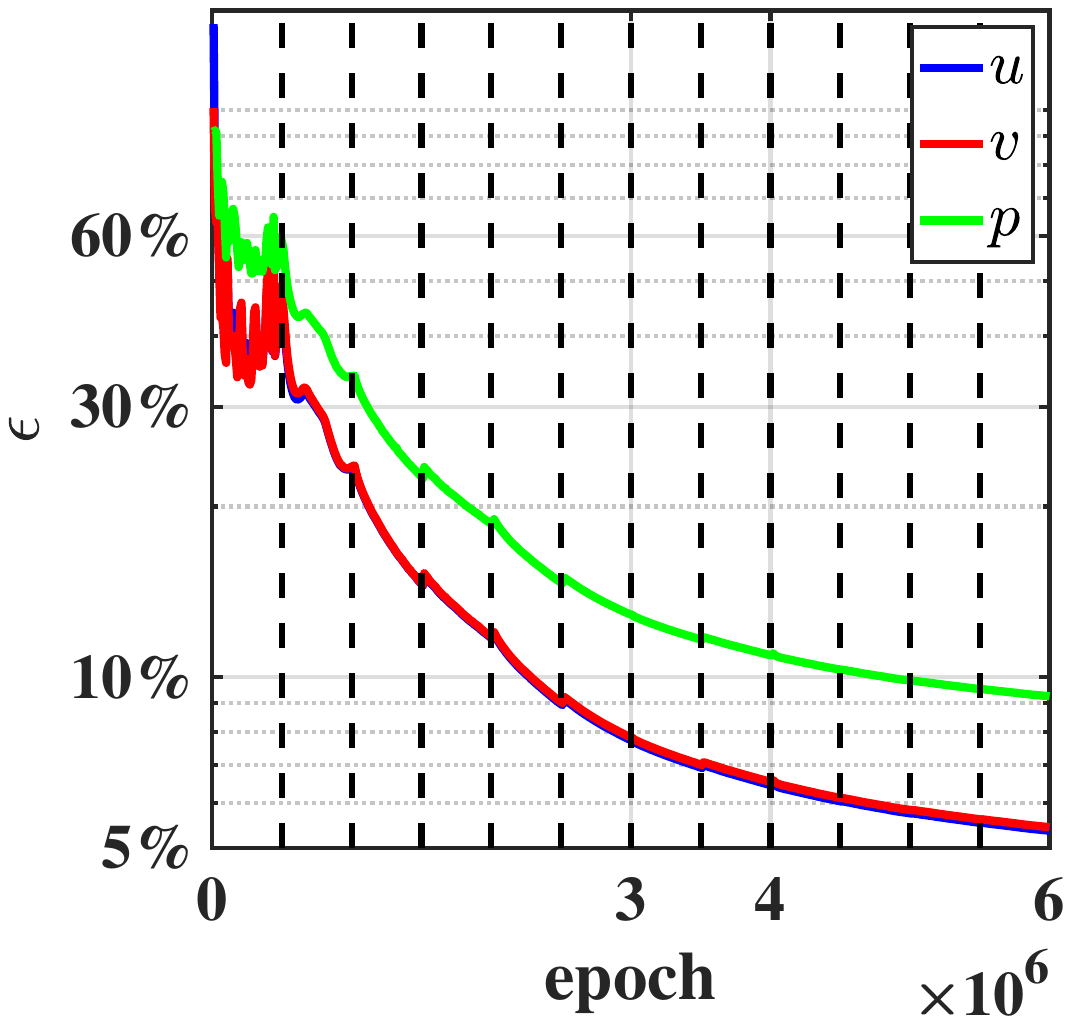}   
   \subcaption{Case H: \\ $Re=5\,000$}
  \end{subfigure} 
  \caption{\textbf{Higher Reynolds number: histories of the loss and relative percent inference error during the training for steady cavity flow.} In both cases, the parameterized eddy viscosity model with two-network structure is used. The spikes of loss at $1.6 \times 10^6$ epoch in Figure (a) and after $3 \times 10^6$ epoch in Figure (b) are due to the restarting of the training. Although the loss function has reached a plateau, the gradient errors of velocity and pressure are still decreasing monotonically, consistent with the information bottleneck theory.}
\label{fig:Loss_hist_high_Re}
\end{figure}%

{\bf Higher Reynolds number}

\begin{figure}
\captionsetup[subfigure]{justification=centering}
\begin{subfigure}{0.31\textwidth}
\includegraphics[width=\textwidth,trim=0 125 0 145,clip]{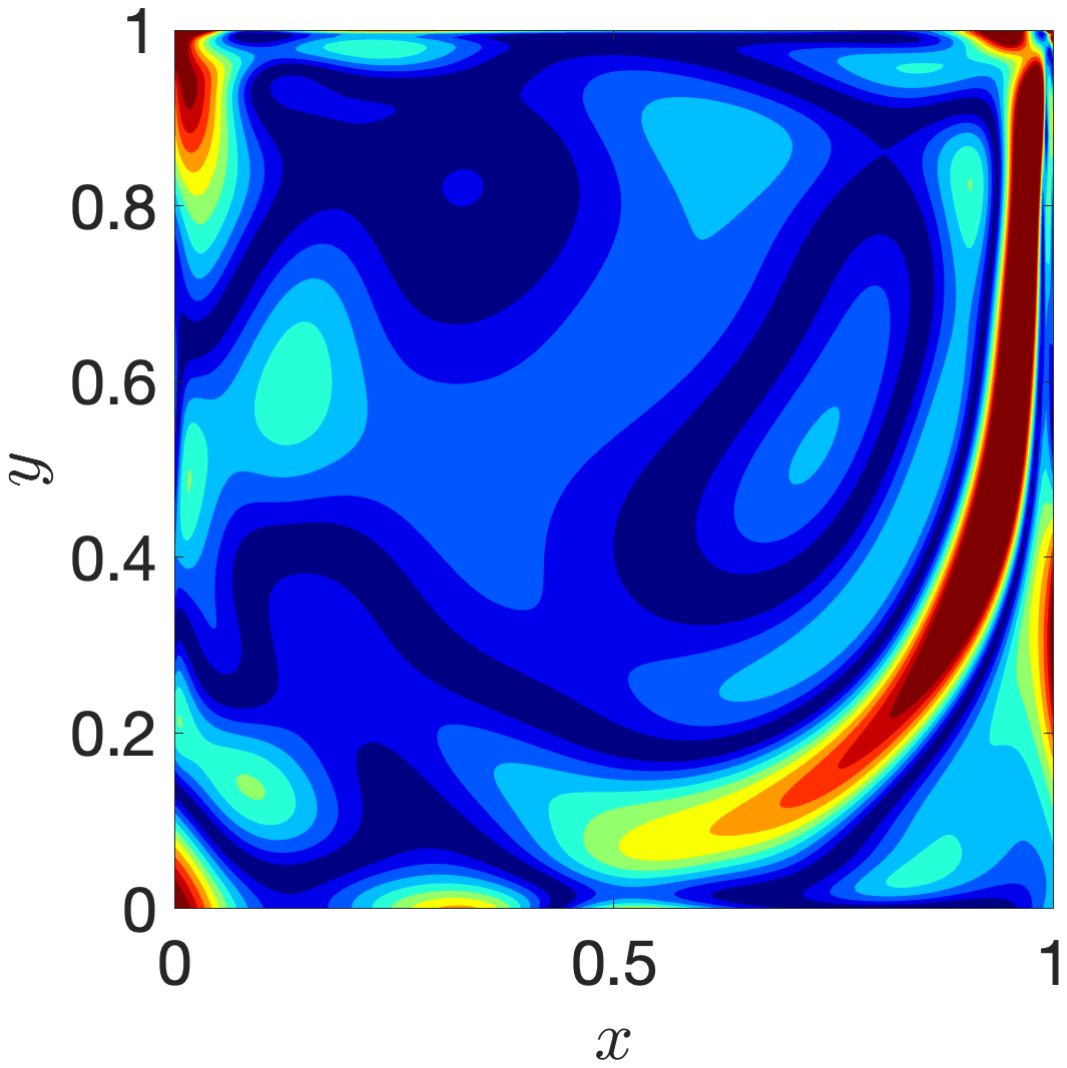}  
\subcaption{$Re=3\,000$; \\
 $\epsilon_u=10.26$, $\epsilon_v=10.42$, $\epsilon_p=14.75$.}
\end{subfigure} 
~
\begin{subfigure}{0.31\textwidth}
\includegraphics[width=\textwidth,trim=0 125 0 145,clip]{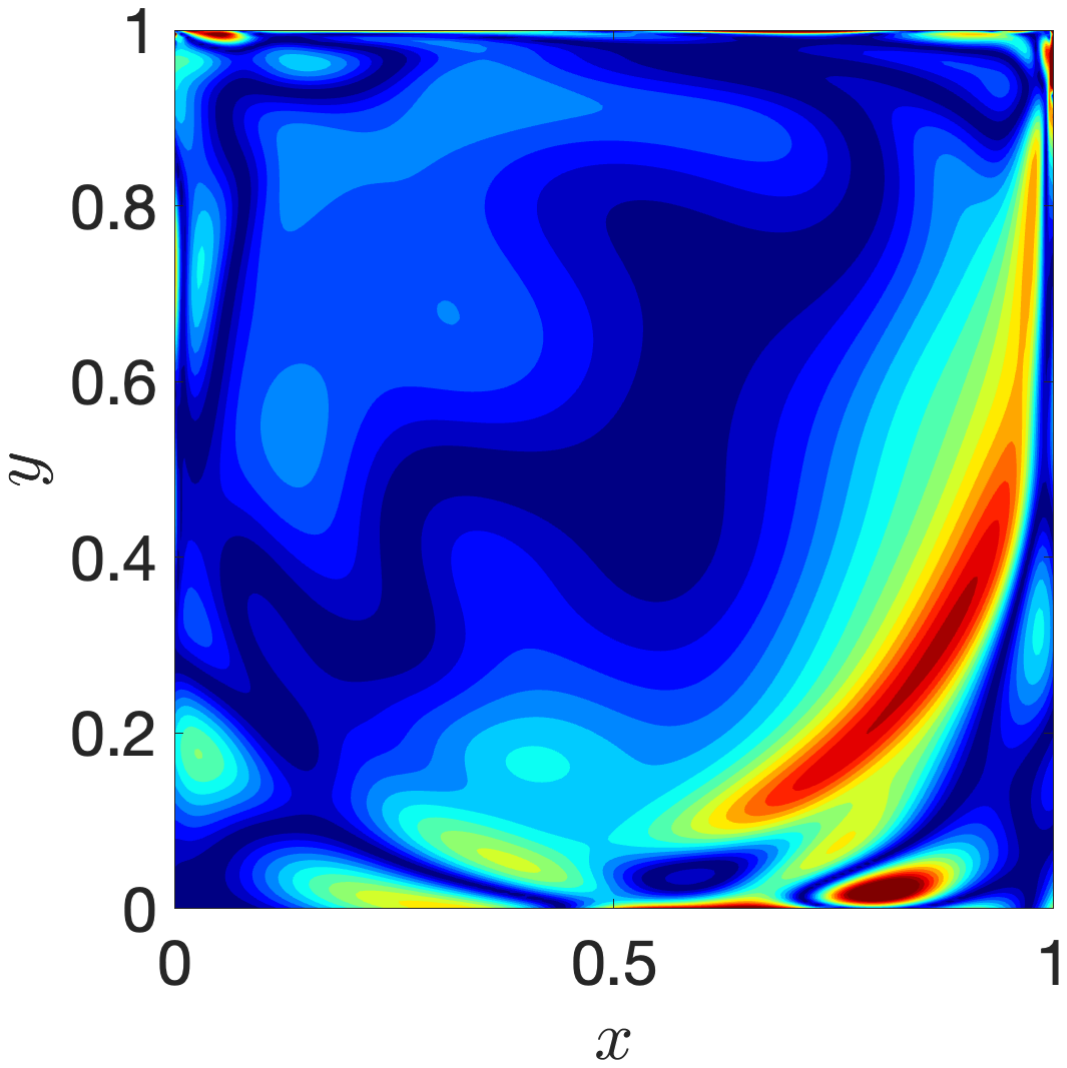}  
  \subcaption{$Re=5\,000$; \\
$\epsilon_u=25.64$, $\epsilon_v=26.19$, $\epsilon_p=46.59$.}
\end{subfigure} 
~
\begin{subfigure}{0.34\textwidth}
\includegraphics[width=\textwidth,trim=0 145 0 145,clip]{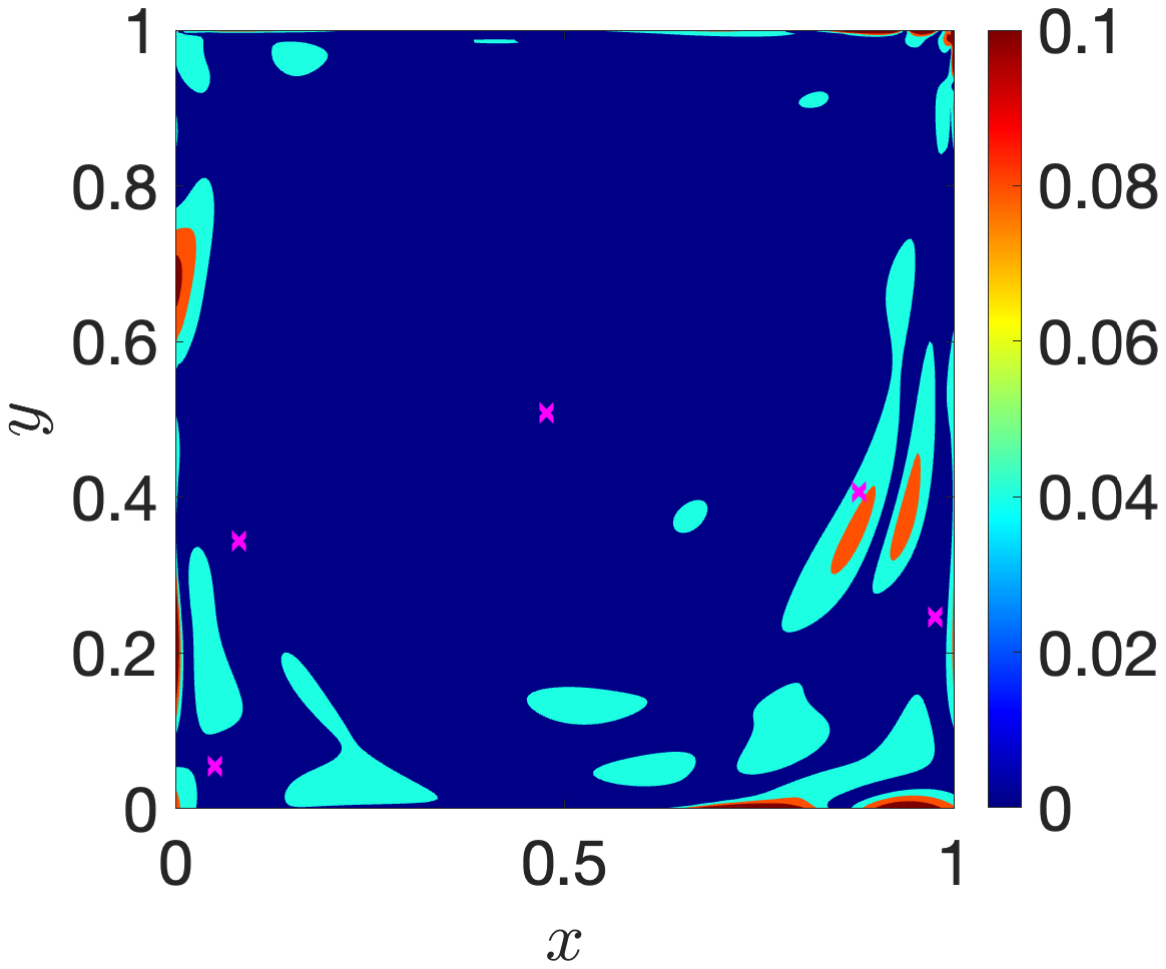}  
  \subcaption{$Re=5\,000$; \\
  $\epsilon_u=3.66$, $\epsilon_v=3.91$, $\epsilon_p=6.78$.}
\end{subfigure} 
~
\caption{\textbf{Transfer learning from lower to higher Reynolds number, showing the normalized eddy viscosity.} Left, transfer from Re=2,000 to Re=3,000; middle, from Re=2,000 to Re=5,000. In both cases, we initialize the network weights by the weights obtained in case E2;  all network weights are trainable; the number of residual points is the same as that in case E2. The eddy viscosity shown in the figure is obtained after training 150,000 epochs with the learning rate (LR) of 1e-5, followed by 100,000 epochs with the learning rate of 2e-6. The right plot shows the normalized eddy viscosity when 5 labeled data points (pink symbols) are available. 
}
~
\label{fig:ev_trans}
\end{figure}

In the previous section, it has been shown that the inference accuracy of the PINNs for Navier-Stokes equation can be improved substantially solely by the eddy viscosity, for the condition that no labeled data is available. However, the ability of the current \emph{single-network parameterized model} for predicting the cavity flow at even higher $Re$ is still limited, e.g., in the case that we employ PINNs for cavity flow at $Re=5,000$, when the network is trained from scratch, the inference RPE never drops below 40\%. In order to improve the PINNs performance for the flow at a higher $Re$, we developed a \emph{two-network parameterized model}, with a separate network for the entropy residual $r$; see Fig. S1 in the supplementary material. The \emph{two-network parameterized model} is applied to infer the cavity flow at $Re=3,000$ and $Re=5,000$ without any labeled data. The loss and error decay history, as well as the distribution of eddy viscosity inferred at the final epoch are plotted in figure \ref{fig:Loss_hist_high_Re}. We see that the \emph{two-network parameterized model} can reduce the RPE of $u,v$ to less than 3\% and 5\%, in the case of $Re=3,000$ and $Re=5,000$, respectively. More interestingly, the MSE loss history in both cases shows a plateau before the error decaying reaching a stagnation. This supports the hypothesis that PINNs training consists of two different stages, namely the fitting and diffusion phases, proposed in recent work by \citep{anagnostopoulos2023residualbased}. Furthermore, the learned eddy viscosity by the \emph{two-network parameterized model} for corresponding $Re$ is presented in the left part of Figure \ref{fig:Loss_hist_high_Re}(a) and (b). We observe that large values of eddy viscosity are mostly located on the four cavity corners, and these value increase with the $Re$. However, the distribution pattern of the eddy viscosity learned by the \emph{two-network parameterized model} is quite different from that by the \emph{single-network  parameterized model}, where the former distribution is organized like a disk, while the latter shows no discerning pattern. In addition to the \emph{two-network parameterized model}, another approach to improve PINNs performance and speed up training for cavity flow at a higher $Re$ is transfer learning. Figure \ref{fig:ev_trans} shows the distribution of the eddy viscosity obtained in transfer learning, where the neural network weights are initialized from the saved values of case B. Nonetheless, as shown in Figures  \ref{fig:ev_trans}(a) and (b), which are for $Re=3,000$ and $Re=5,000$, respectively, the eddy viscosity from transfer learning exhibits a prominent value along the border of the primary vortex, and the final RPE values for both cases are relatively large. In contrast, using 5 different labeled data as shown on the right plot at $Re=5,000$ leads to substantially smaller errors.
\section{Discussion}
In the last 50 years there has been great progress in direct numerical simulation (DNS) of fluid flows in simple and complex geometries but the value of Reynolds number at which we can resolve all energetic scales is still relatively low. Hence, large eddy simulation (LES) that uses explicitly an eddy viscosity model has been employed to increase the range of feasible Reynolds numbers. Physics-informed neural networks (PINNs) is a relative new method that has shown promise in blending physical laws and available data smoothly. However, its use in predicting solution of fluid flows at high Reynolds number is problematic because the loss landscapes become increasingly rough as the multiscale features of the flow increase. Moreover, for the classic problem of the lid-driven cavity flow that we studied herein, we observed that we obtained another class of solutions to the Navier-Stokes equations, which are associated with a local minimum. This class of solutions cannot be captured by DNS, which targets the most energetic solution. Interestingly, by employing an LES formulation in PINNs and fine tune the parametric viscosity we can obtain unique solutions at different Reynolds numbers (from 2,000 to 5,000), which are very close to the DNS solutions. The exact form of the eddy viscosity may not even be important as we have seen with the two different approaches we pursued, namely a parameterized model or a viscosity function learned directly by a neural network.

If instead of the eddy viscosity, we use labeled data at scattered points, even one single point measurement, we still obtain unique solutions close to the DNS solutions, which are physically realizable.  This can be explained by examining the loss functions landscapes that seem to become smoother either by injecting eddy viscosity or labeled data in the PINNs formulation. These new findings, in turn, could guide us on how to formulate PINN methods in the future for studying high Reynolds number flows that are currently out of reach of the classical DNS or LES studies. 

With regards to multiplicity of solutions that we encountered here, theoretical work suggests that 
for steady state flow problems there is a unique steady state solution provided  that the viscosity is large and the boundary data is small, i.e., small Reynolds number. However, when the viscosity gets smaller (or the boundary data gets larger), we should expect some kind of bifurcation, which translates into the co-existence of more than one steady state. Of course some of these steady states need not be stable, 
as we realized in the current study where only one of the two solutions matched the DNS solution corresponding to a global minimum of the loss landscape.


\appendix

\section{ev-NSFnet network structure}\label{appA}

\begin{figure}
\centering
  \begin{subfigure}{0.475\textwidth}
\raisebox{1.5cm}{
\includegraphics[width=0.9\textwidth,trim=0 0 0 0,clip]{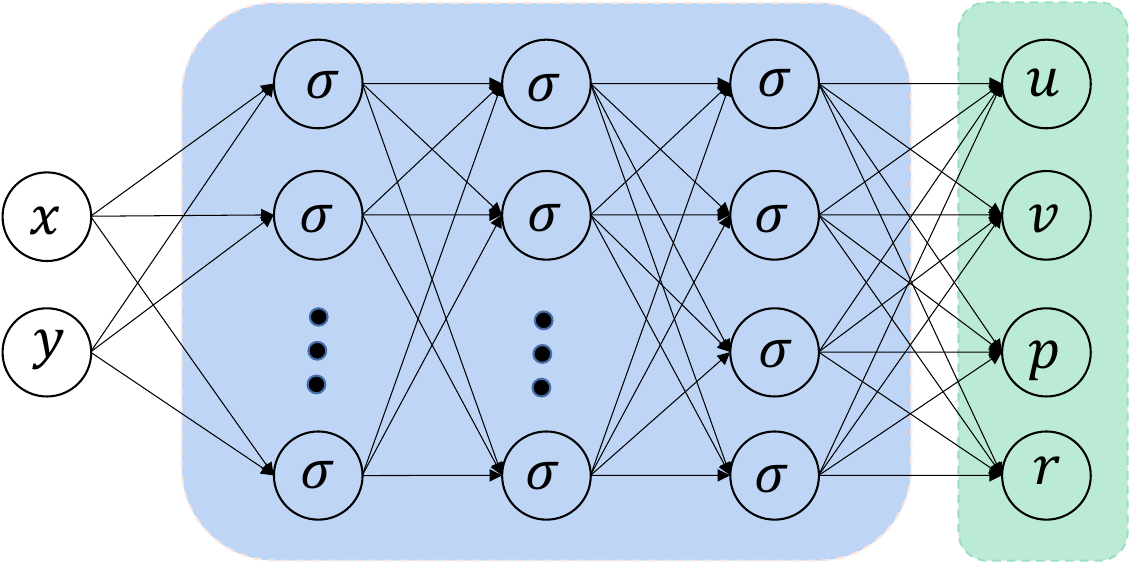}}%
\subcaption{\emph{one-network}}
\end{subfigure} 
~
\centering
 \begin{subfigure}{0.475\textwidth}
   \includegraphics[width=0.9\textwidth,trim=0 0 0 0,clip]{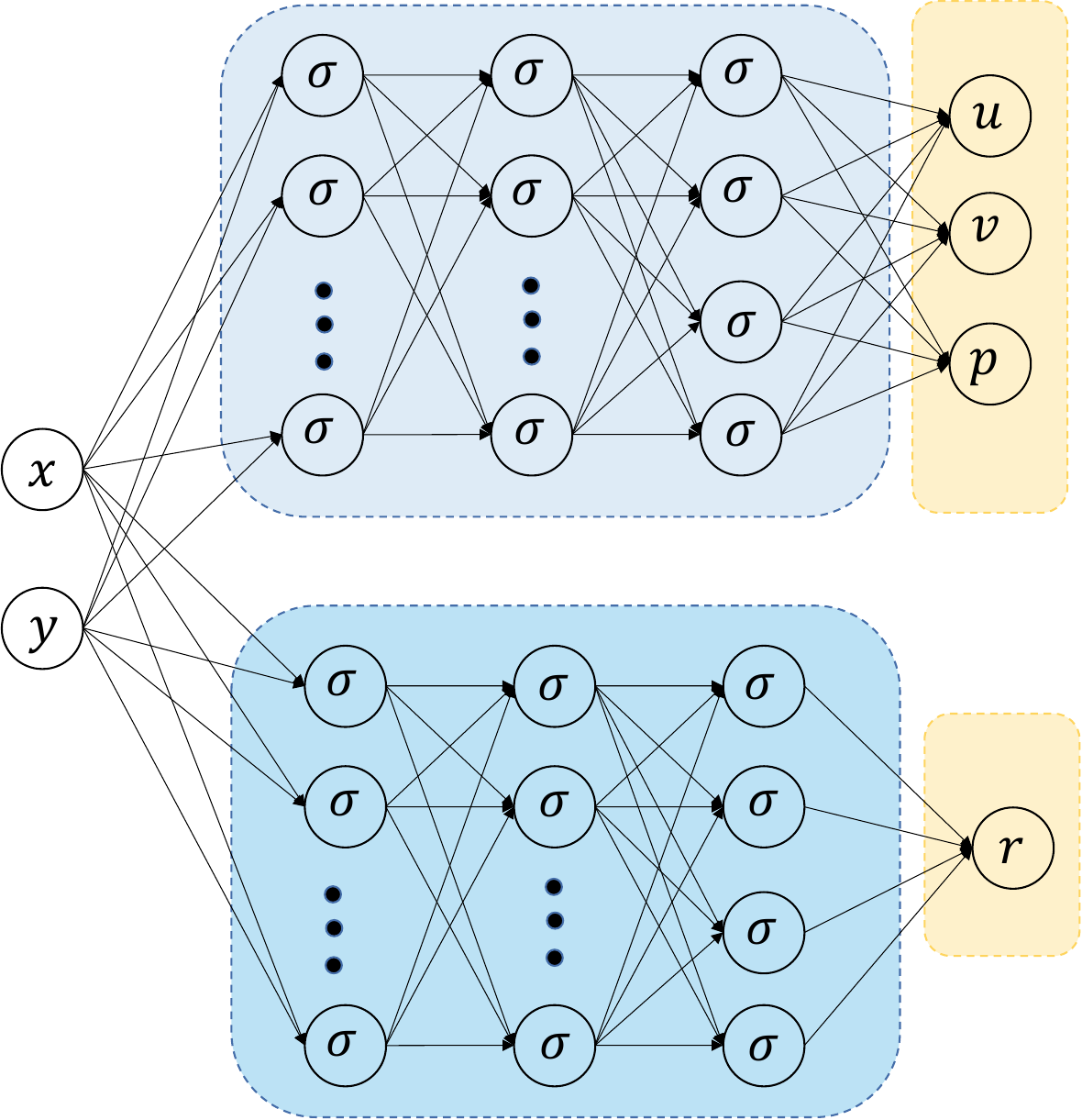}
    \subcaption{\emph{two-network}}
  \end{subfigure} 
  \caption{Network architectures of ev-NSFnet using the \emph{parameterized eddy viscosity model} for cavity flow at high $Re$: $x$ and $y$ are the coordinates, $u$, $v$, $p$ are the velocity components and pressure, respectively, and $r$ is prediction of equation residual, which is defined in Eq. 10 in the main text; $\sigma$ denotes the activation function.}
\label{fig:net_struct}
\end{figure}

In this section, we provide details of the ev-NSFnet with \emph{parameterized eddy viscosity model} for predicting the 2D steady cavity flow at $Re \ge 2,000$, without using any labeled data within the computational domain. As shown in Fig. \ref{fig:net_struct}, different from the original NSFnet developed in \citep{JIN2021109951}, ev-NSFnet has a new output variable $r$, which is the  equation's entropy residual defined by Eq. 10 in the main text. In particular,  Figs. \ref{fig:net_struct} (a) and (b) show the the \emph{single-network} and \emph{two-network} model, respectively. The \emph{single-network} consists of $6 \times 80$ hidden neurons, while the \emph{two-network} model is made of $4 \times 120$ hidden neurons for $u$, $v$ and $p$, and $4 \times 40$ hidden neurons for $r$. Our extensive ev-NSFnet applications show that the \emph{single-network} model works well when $Re \le 3,000$, while the \emph{two-network} model works for all $Re$, despite the additional complexity in terms of implementation for the latter. 

It is worth noting that in the training of ev-NSFnet, the entropy viscosity $\nu_E$ at $n^{\text{th}}$ epoch, is computed from $r$ obtained at $(n-1)^{\text{th}}$ epoch. For more details about the implementation, please refer to our GitHub pape \citep{NSFnet_github}.


\section{Effect of $\alpha$ (ev model) and $N_e$ (residue points)} \label{appB}

In this section, we present a study on the effect of $\alpha$ of the \emph{one-network entropy-viscosity parameterized} model. Note that similarly to the implementation of the entropy viscosity method in numerical simulation \citep{wang2019jfm}, $\beta$ is the parameter that sets the upper limit of the entropy viscosity, while $\alpha$ is the actual parameter that controls the value of entropy viscosity. Therefore, in the current study, a fixed value $\beta=5$ has been used, and the range $\alpha \in [0,0.5]$ has been investigated. In addition, the impact of the number of residual points ($N_e$) on the relative percentage error (RPE) of the solution has been studied. Note that, due to the randomness of neural network training, the case of exactly the same parameters has been performed 5 times. Moreover, in order to give a fair comparison, the same number of training epochs and same learning rate $l_r$ have been employed. In particular, as shown in Table \ref{Re2K_lr}, for the case using the $single-network$ model, the entire training process is divided into 5 stages, while in the case of using the $two-network$ model, the training consists of 6 stages.  Nonetheless, before proceeding to the results, it is worth noting that the residual points in the computational domain are generated by the Latin hypercube sampling (LHS) scheme.

Table \ref{rpe_const_lr} presents the inferred RPE for cavity flow at $Re=2,000$ and $Re=3,000$, using different $\alpha$ and $N_e$. It could be observed that $\alpha=0.03$ produces the most accurate result, with the mean RPE in velocity falling below 5\%. The mean RPE in velocity  of $\alpha=0.01$ is around 40\%, while that of $\alpha=0.05$ is around 28\%. Note that although the accuracy is different in eacch case, the ev-NSFnet gives rise to \emph{class 2} type of solution only. This pattern is similar to that produced by DNS, see explanation in the main text of this paper. Note that for any case in Table \ref{rpe_const_lr}, the RPE can be reduced further, if additional  training is performed. When $\alpha=0.03$ is applied to the cavity flow at $Re=3,000$, the mean RPE in velocity is less than 7\%.  Furthermore, it could be observed that the number of residual points has a notable impact on the RPE. $N_e=4 \times 10^4$ leads to accurate prediction at $Re=2,000$, but when it comes to $Re=3,000$, the mean RPE is increased to around 20\%. However, as the $N_e$ is increased to $6 \times 10^4$, the mean RPE falls below 7\%. Note that the mean RPE rises beyond 12\%, with $N_e$ further increasing to $8 \times 10^4$, but the variation of the RPE among the 5 simulations using $N_e=8 \times 10^4$ is smaller than that using $N_e=6 \times 10^4$.

\begin{figure*}
\centering
  \begin{subfigure}{0.475\textwidth}
\includegraphics[width=\textwidth,trim=0 100 0 0,clip]{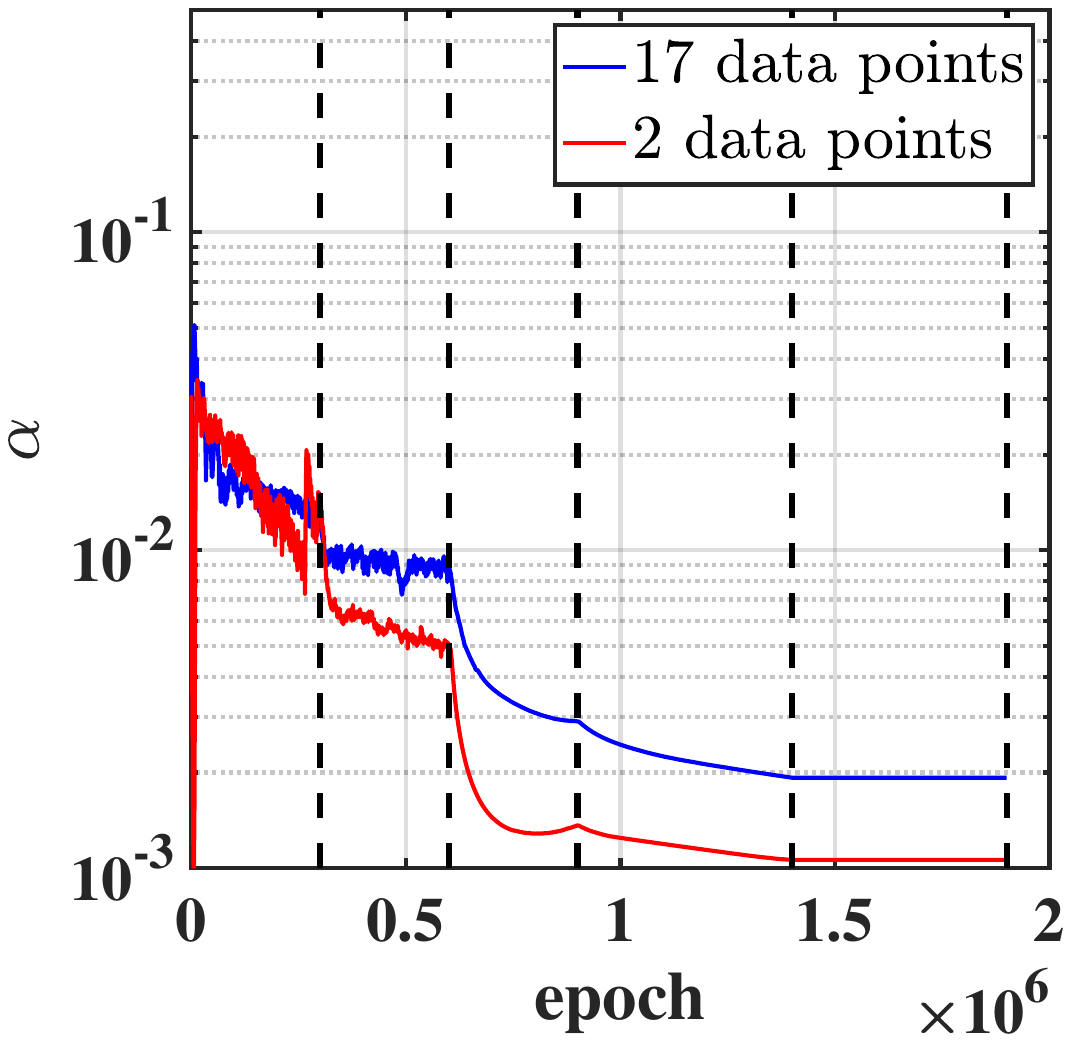}%
\end{subfigure} 
~
\centering
 \begin{subfigure}{0.475\textwidth}
   \includegraphics[width=\textwidth,trim=0 100 0 0,clip]{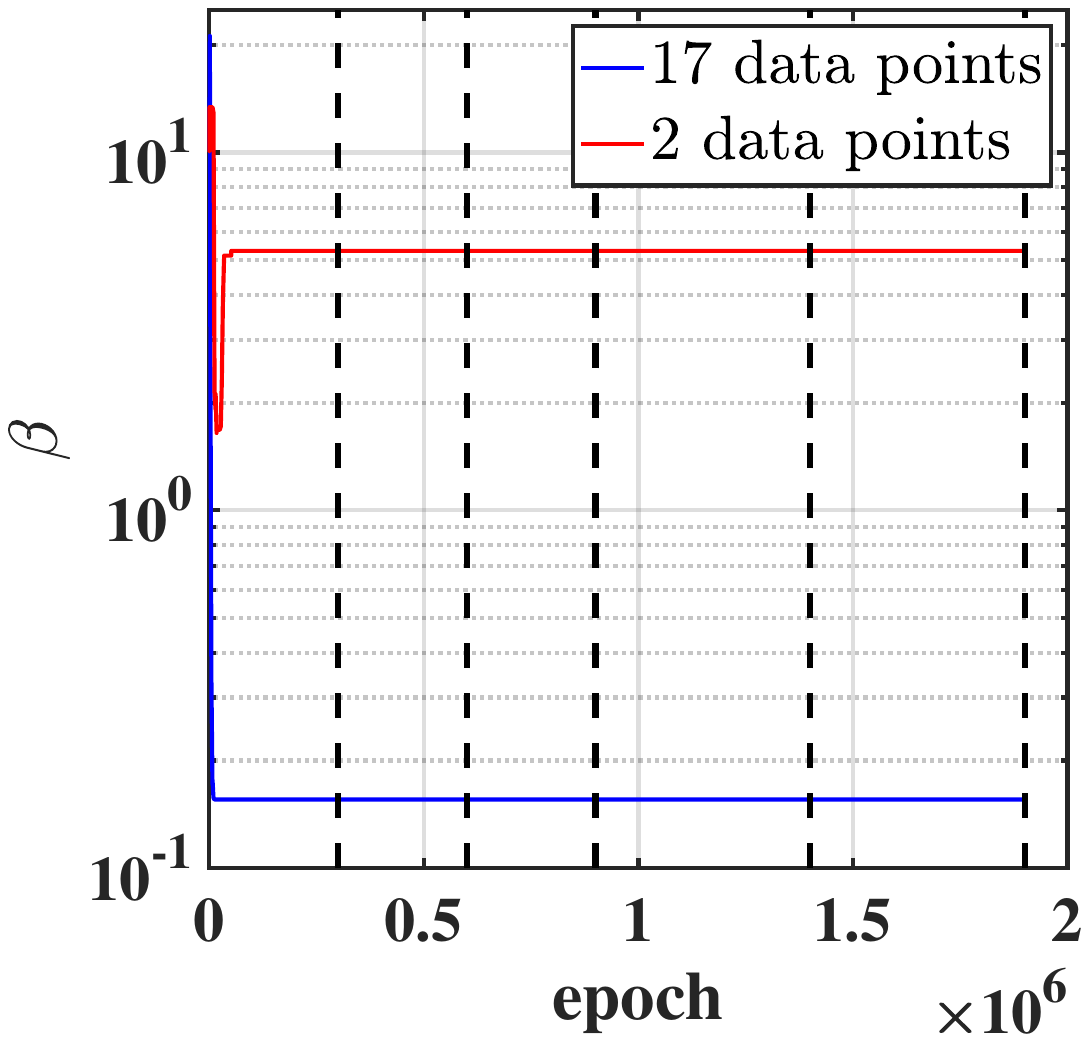}
  \end{subfigure} 
  \caption{Parameterized eddy viscosity: inferred values of $\alpha$, $\beta$ by ev-NSFnet for the inverse problem of cavity flow at $Re=2\,000$, where the additional data points for the velocity are at the center lines of the domain. Note that `2 data points' and `17 data points' mean $\alpha$ and $\beta$ are inferred by ev-NSFnet with 2 and 7 labeled data points, respectively.}
\label{fig:alpha_beta}
\end{figure*}

The ev-NSFnet prediction can be further improved by using a varying $\alpha$, in a similar fashion as the learning rate. Table \ref{Re2K_lr} lists the value of $\alpha$ used in each training stage. Actually, the $\alpha$ can be learned from data. To this end, the PINNs inverse problem can be setup as follows: $\alpha$ or $\beta$ is unknown variable, but there are a few velocity measurements in the domain. In the case that 2 and 17 measurement points are available,  the solutions of the PINNs inverses problem are given in Figure \ref{fig:alpha_beta}. It could be observed that in both cases the learned $\beta$ quickly reaches a constant value after a few epochs. In particular, $\beta$ approaches $\beta \approx 5$ in the case of 2 measurement points, which indicates $\nu_E \le 5 \nu$. Furthermore, in both cases, $\alpha$ is decreasing with the training epochs, e.g., as shown by the red curve of Figure \ref{fig:alpha_beta} (a), $\alpha \approx 0.03$ in the very beginning of the training, and it decreases to 0.001 in the final stage of training. Keeping in mind that $\alpha$ is decreasing with training epochs and $\beta$ reaches a constant, in the current PINNs forward problem, $\beta=5$ has been used, while the $\alpha$ used in each training stage is listed in Table \ref{Re2K_lr}. Table \ref{rpe_variable_lr} presents the RPE at $Re=2,000$, $Re=3,000$ and $Re=5,000$, when $\alpha$ is a variable. It could be observed that the RPE at $Re=2\,000$ and $Re=3,000$ are notably smaller than that of constant $\alpha$. Notably, the inferred RPE in velocity at $Re=5,000$ can be reduced to within 7\%.

To sum up, in the scenario of no labeled data, for the ev-NSFnet with \emph{parameterized eddy viscosity model}, there exists an optimal $\alpha$ that can lead to most accurate prediction. Moreover, in order to obtain accurate PINNs solution, a sufficiently large $N_e$ is required. 
\section{Additional results at $Re=5,000$}\label{appC}
\begin{figure}[h]
\centering
  \begin{subfigure}{0.475\textwidth}
\includegraphics[width=\textwidth,trim=150 20 150 20,clip]{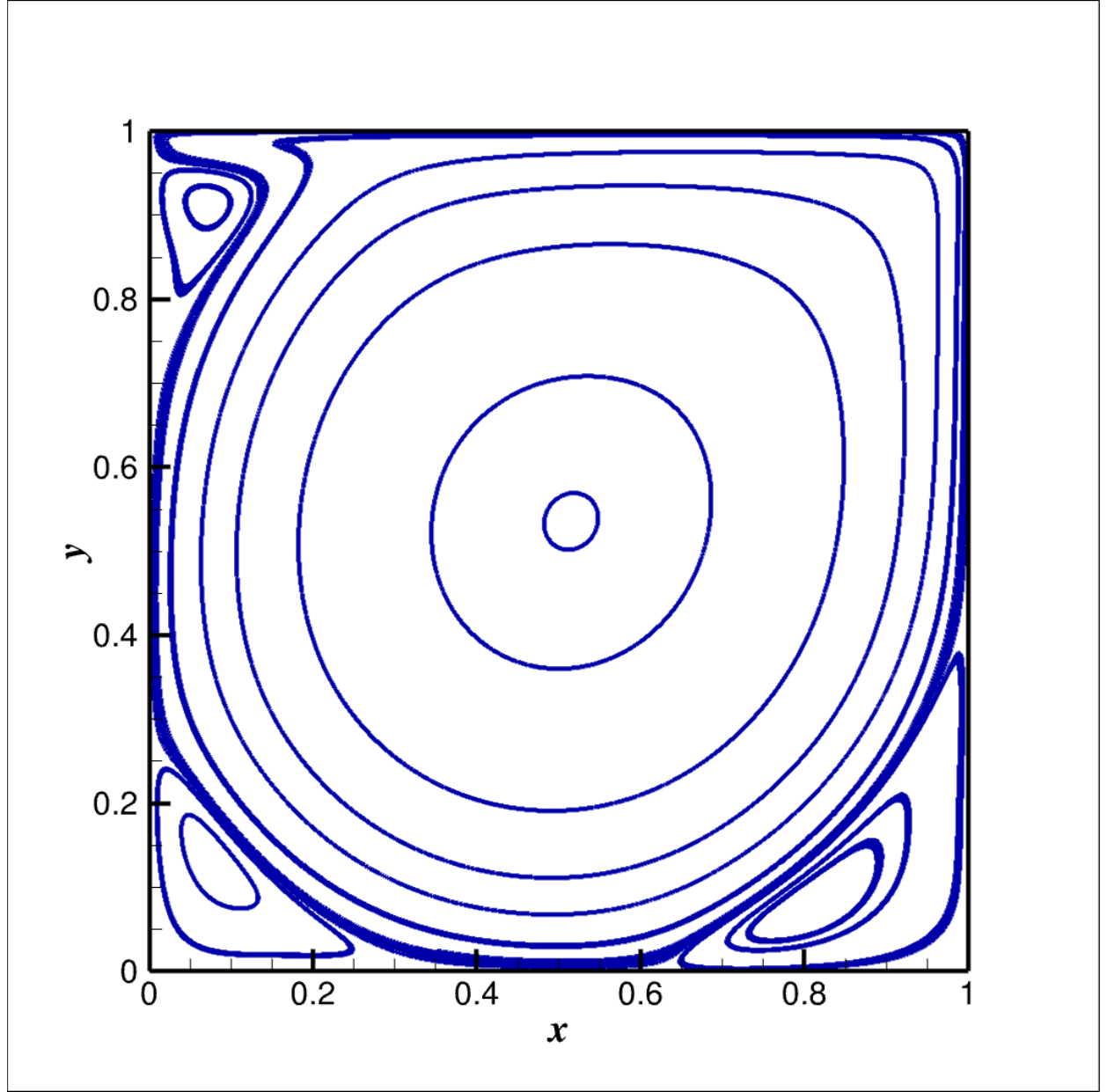}%
\subcaption{CFD}
\end{subfigure} 
~
\centering
 \begin{subfigure}{0.475\textwidth}
   \includegraphics[width=\textwidth,trim=150 20 150 20,clip]{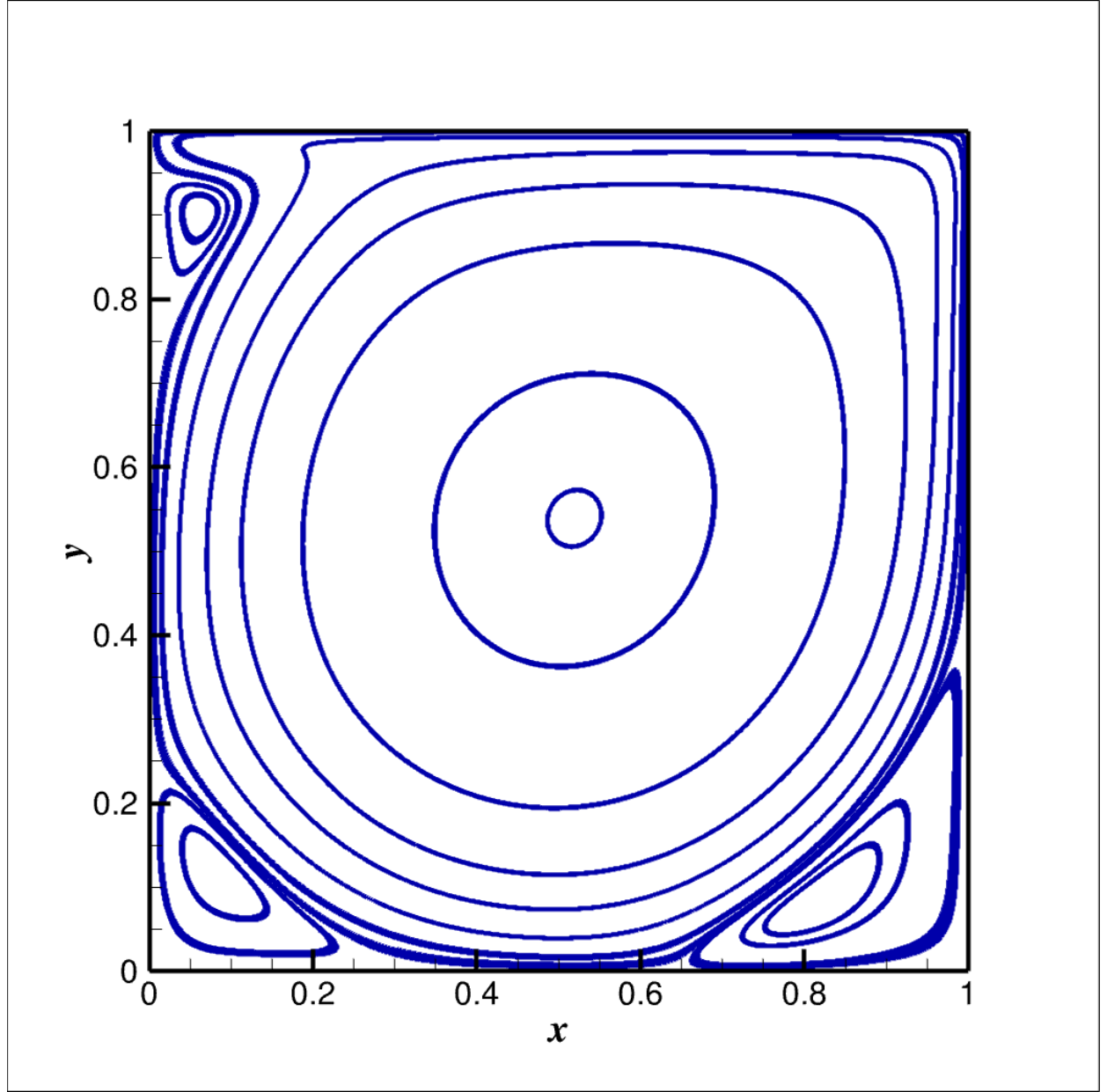}
   \subcaption{ev-NSFnet} 
  \end{subfigure} 
  \caption{Comparison between high-order CFD and ev-NSFnet (Case D1): streamlines, $Re=5,000$. }
\label{fig:Re5K_streamline}
\end{figure}

\begin{figure}[h]
\centering
  \begin{subfigure}{0.475\textwidth}
\includegraphics[width=\textwidth,trim=0 100 0 100,clip]{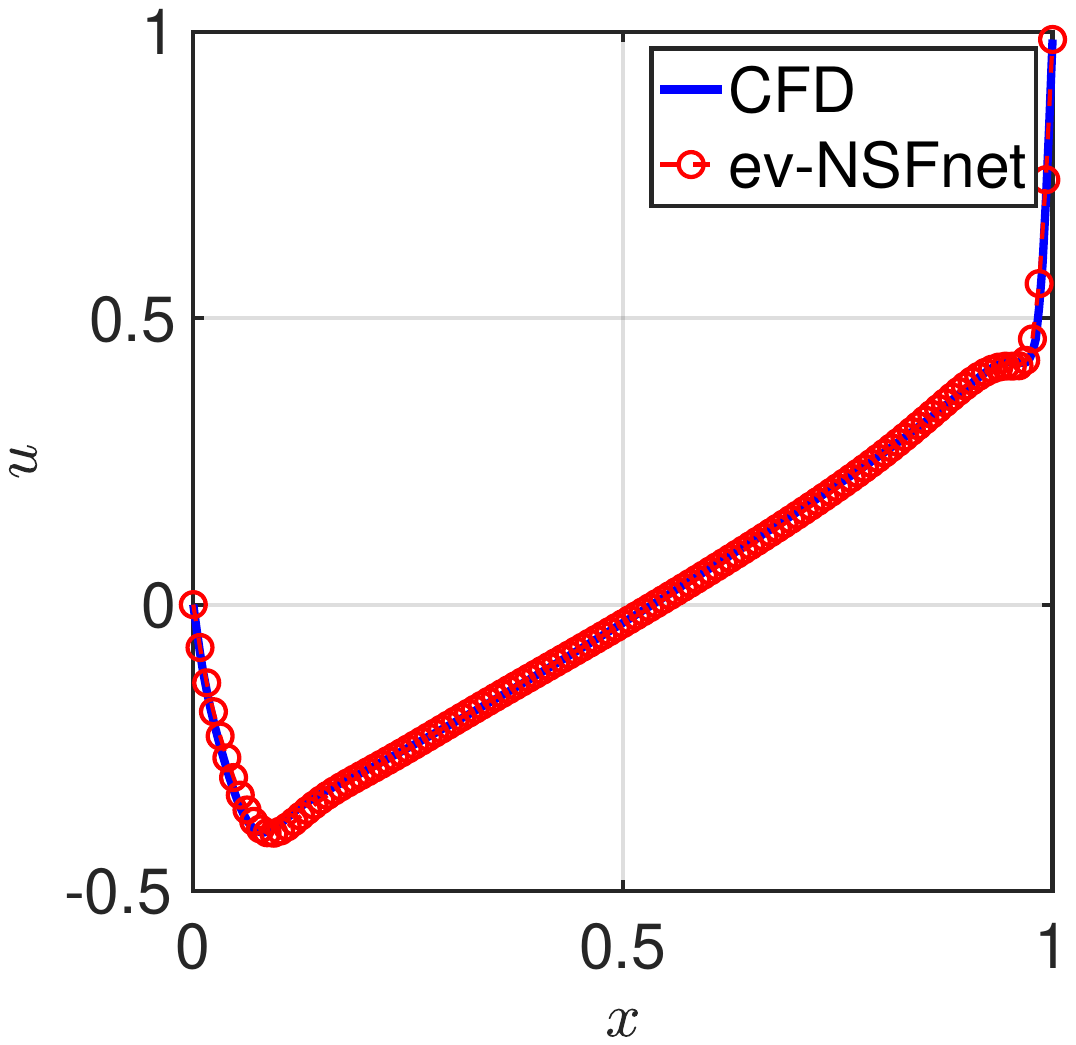}%

\end{subfigure} 
~
\centering
 \begin{subfigure}{0.475\textwidth}
   \includegraphics[width=\textwidth,trim=0 100 0 100,clip]{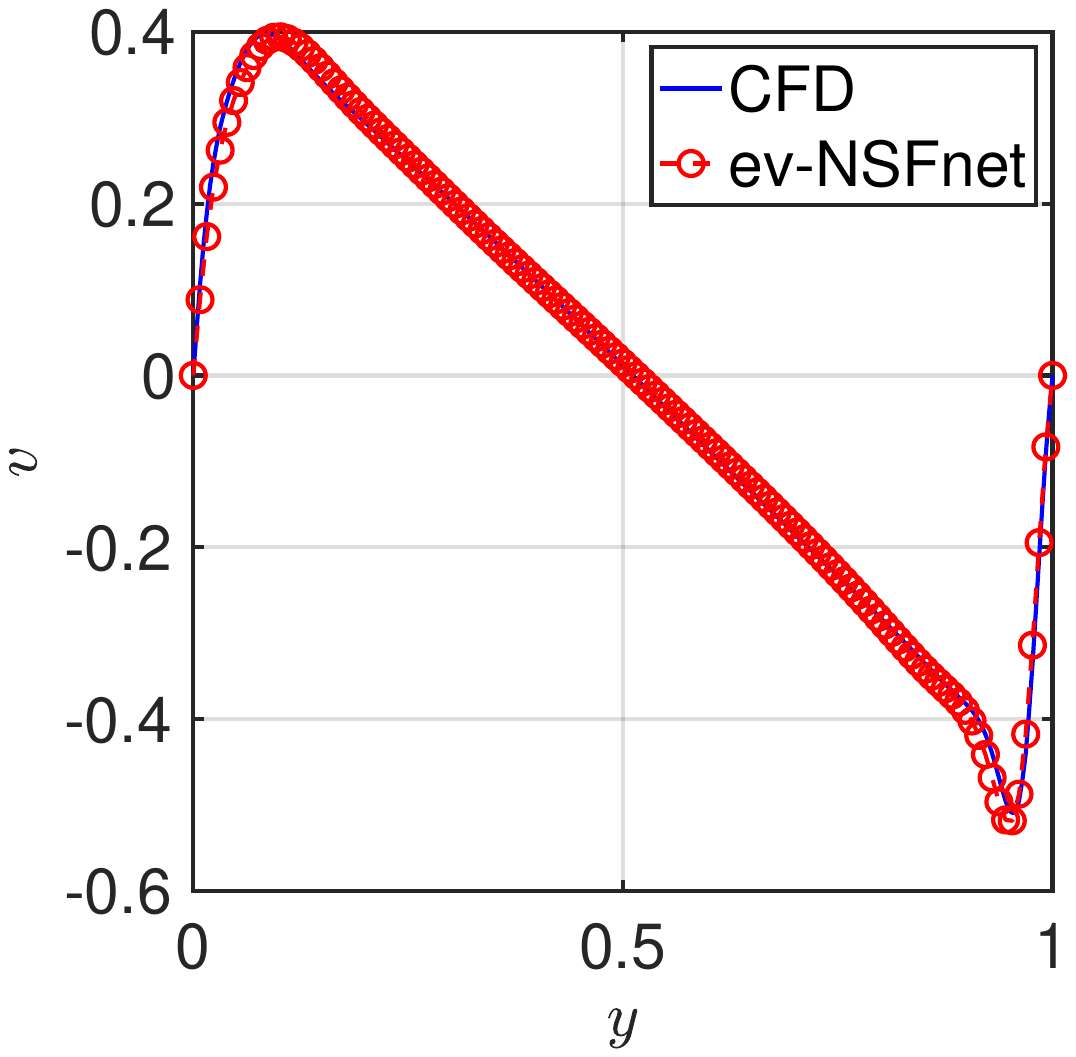}
    
  \end{subfigure} 
  \caption{Comparison between high-order CFD and ev-NSFnet (Case D1): distribution of $u$ and $v$ along $y=0.5$ and $y=0.5$, respectively, at $Re=5,000$.}
\label{fig:Re5K_uv}
\end{figure}

In this section, the comparison between ev-NSFnet and high-order CFD simulation for the cavity flow at $Re=5,000$ is given. Note that here the ev-NSFnet result is taken from case D1 presented in Table \ref{rpe_variable_lr}. Figure \ref{fig:Re5K_streamline} presents the quality comparison on the streamlines. It could be observed that the ev-NSFnet captures the primary vortex as well the secondary vortices in the corners very accurately. Figure \ref{fig:Re5K_uv} (a) and (b) plot the quantitative comparison on the velocity profiles along the cavity center lines $y=0.5$ and $x=0.5$, respectively. Again, it can be observed that the prediction by ev-NSFnet agrees with the CFD very well.

\renewcommand{\arraystretch}{1.0}
\begin{table}
\caption{Sensitivity of ev-NSFnet result to parameter $\alpha$ and number of residual points $N_e$. $\epsilon$ is the relative percent error (RPE). $\beta=5$ is used in all cases. No labeled data is used within the computational domain. Note that the error is defined as $\epsilon_{\phi}=\frac{\| \hat{\phi}-\phi\|_2}{ \|\phi\|_2} \times 100$, where $\phi$ denotes the reference data, $\hat{\phi}$ is the PINNs regressed value, computed on the uniform $256\times 256$ mesh. }
\begin{center}
\begin{adjustbox}{width=0.95\textwidth}
\begin{NiceTabular}{|ccccccccccccccccccc}
\toprule
\multirow{5}{2em}{\textcolor{blue}{\textbf{Case}}} 
& \multicolumn{3}{|c|}{\textcolor{blue}{\textbf{I}}} 
& \multicolumn{3}{|c|}{\textcolor{blue}{\textbf{E}}} 
& \multicolumn{3}{|c|}{\textcolor{blue}{\textbf{J}}} 
& \multicolumn{3}{|c|}{\textcolor{blue}{\textbf{K}}} 
& \multicolumn{3}{|c|}{\textcolor{blue}{\textbf{L}}} 
& \multicolumn{3}{|c|}{\textcolor{blue}{\textbf{M}}}
\\
\cline{2-19}
& \multicolumn{9}{|c|}{$Re=2000$} 
& \multicolumn{9}{|c|}{$Re=3000$} \\
\cline{2-19}
& \multicolumn{9}{|c|}{$N_e=4 \times 10^4$} 
& \multicolumn{3}{|c|}{$N_e=4 \times 10^4$} 
& \multicolumn{3}{|c|}{$N_e=6 \times 10^4$} 
& \multicolumn{3}{|c|}{$N_e=8 \times 10^4$} 
\\
\cline{2-19}
&\multicolumn{3}{|c|}{\makecell[c]{$\alpha=0.01$}}
&\multicolumn{3}{|c|}{\makecell[c]{$\alpha=0.03$}}
&\multicolumn{3}{|c|}{\makecell[c]{$\alpha=0.05$}}
&\multicolumn{9}{|c|}{\makecell[c]{$\alpha=0.03$}}
\\
\cline{2-19}
&\multicolumn{3}{|c|}{\makecell[c]{$\epsilon_u\quad\, \epsilon_v \quad\, \epsilon_p$}}
&\multicolumn{3}{|c|}{\makecell[c]{$\epsilon_u\quad\, \epsilon_v \quad\, \epsilon_p$}}
&\multicolumn{3}{|c|}{\makecell[c]{$\epsilon_u\quad\, \epsilon_v \quad\, \epsilon_p$}}
&\multicolumn{3}{|c|}{\makecell[c]{$\epsilon_u\quad\, \epsilon_v \quad\, \epsilon_p$}}
&\multicolumn{3}{|c|}{\makecell[c]{$\epsilon_u\quad\, \epsilon_v \quad\, \epsilon_p$}}
&\multicolumn{3}{|c|}{\makecell[c]{$\epsilon_u\quad\, \epsilon_v \quad\, \epsilon_p$}}
\\
\midrule
\textcolor{blue}{\textbf{1}}
&\multicolumn{3}{|c|}{\makecell[c]{36.8\, 38.0\, 58.9}}
&\multicolumn{3}{|c|}{\makecell[c]{3.7\,  3.8\,  7.5}}
&\multicolumn{3}{|c|}{\makecell[c]{33.8\, 35.6\, 55.0}}
&\multicolumn{3}{|c|}{\makecell[c]{21.7\, 22.6\, 40.0}} 
&\multicolumn{3}{|c|}{\makecell[c]{2.9\, 2.8\, 4.8}} 
&\multicolumn{3}{|c|}{\makecell[c]{11.8\, 11.9\, 22.7}}
\\
\midrule
\textcolor{blue}{\textbf{2}}
&\multicolumn{3}{|c|}{\makecell[c]{37.4\, 38.3\, 58.8}}
&\multicolumn{3}{|c|}{\makecell[c]{3.6\,  3.7\,  7.2}}
&\multicolumn{3}{|c|}{\makecell[c]{16.4\, 13.7\, 14.0}}
&\multicolumn{3}{|c|}{\makecell[c]{5.3\, 5.5\, 10.8}} 
&\multicolumn{3}{|c|}{\makecell[c]{10.1\, 10.3\, 19.8}} 
&\multicolumn{3}{|c|}{\makecell[c]{10.9\, 11.0\, 21.1}} 
\\
\midrule
\textcolor{blue}{\textbf{3}}
&\multicolumn{3}{|c|}{\makecell[c]{30.8\, 31.7\, 51.8}}
&\multicolumn{3}{|c|}{\makecell[c]{4.2\, 4.5\, 8.9}}
&\multicolumn{3}{|c|}{\makecell[c]{22.6\, 20.7\, 25.0}}
&\multicolumn{3}{|c|}{\makecell[c]{35.8\, 36.8\,57.9}} 
&\multicolumn{3}{|c|}{\makecell[c]{12.8\, 13.0\, 24.5}} 
&\multicolumn{3}{|c|}{\makecell[c]{11.0\, 11.1\, 21.2}}
\\
\midrule
\textcolor{blue}{\textbf{4}}
&\multicolumn{3}{|c|}{\makecell[c]{34.1\, 34.9\, 55.2}}
&\multicolumn{3}{|c|}{\makecell[c]{7.6\, 7.8\, 14.9}}
&\multicolumn{3}{|c|}{\makecell[c]{11.4\, 9.9\, 10.1}}
&\multicolumn{3}{|c|}{\makecell[c]{16.8\, 17.2\, 31.4}}
&\multicolumn{3}{|c|}{\makecell[c]{13.9\, 14.1\, 26.5}} 
&\multicolumn{3}{|c|}{\makecell[c]{8.4\, 8.6\, 16.7}} 
\\
\midrule
\textcolor{blue}{\textbf{5}}
&\multicolumn{3}{|c|}{\makecell[c]{41.4\, 42.5\, 63.5}}
&\multicolumn{3}{|c|}{\makecell[c]{4.5\, 4.7\, 9.1}}
&\multicolumn{3}{|c|}{\makecell[c]{21.6\, 21.9\, 38.4}}
&\multicolumn{3}{|c|}{\makecell[c]{15.6\, 16.1\, 29.6}} 
&\multicolumn{3}{|c|}{\makecell[c]{10.2\, 10.4\, 19.9}}
&\multicolumn{3}{|c|}{\makecell[c]{13.1\, 13.2\, 25.7}}
\\
\midrule
\textcolor{red}{\textbf{mean}} 
&\multicolumn{3}{|c|}{\makecell[c]{\textcolor{red}{39.1\, 40.3\, 61.2}}}	
&\multicolumn{3}{|c|}{\makecell[c]{\textcolor{red}{4.1\,  4.3\,	8.3}}}
&\multicolumn{3}{|c|}{\makecell[c]{\textcolor{red}{27.7\, 28.8\, 46.7}}}
&\multicolumn{3}{|c|}{\makecell[c]{\textcolor{red}{18.7\, 19.4\, 34.8}}}
&\multicolumn{3}{|c|}{\makecell[c]{\textcolor{red}{6.55\, 6.6\,	 12.35}}}
&\multicolumn{3}{|c|}{\makecell[c]{\textcolor{red}{12.45\,12.6\, 24.2}}}
 \\
\bottomrule
\end{NiceTabular}
\end{adjustbox}
\end{center}
\label{rpe_const_lr}
\end{table}
\renewcommand{\arraystretch}{1.0}
\begin{table}
\caption{Relative percent errors ($\epsilon$) of PINNs for cavity flow at high $Re$. No labeled data is used within the computational domain. $\epsilon$ is the relative percentage error, whose definition is given in caption of Table \ref{rpe_const_lr}.}
\begin{center}
\begin{adjustbox}{width=0.9\textwidth}
\begin{NiceTabular}{|cccccccccccccccc}
\toprule
\multirow{6}{2em}{\textcolor{blue}{\textbf{Case}}} 
& \multicolumn{3}{|c|}{\textcolor{blue}{\textbf{A}}} 
& \multicolumn{3}{|c|}{\textcolor{blue}{\textbf{B}}} 
& \multicolumn{3}{|c|}{\textcolor{blue}{\textbf{C}}} 
& \multicolumn{3}{|c|}{\textcolor{blue}{\textbf{G}}} 
& \multicolumn{3}{|c|}
{\textcolor{blue}{\textbf{H}}} 
\\
\cline{2-16}
& \multicolumn{3}{|c|}{NSFnet} 
& \multicolumn{9}{|c|}{ev-NSFnet} \\
\cline{2-16}
& \multicolumn{6}{|c|}{$Re=2000$} 
& \multicolumn{6}{|c|}{$Re=3000$}
& \multicolumn{3}{|c|}{$Re=5000$} \\
\cline{2-16}
& \multicolumn{6}{|c|}{$N=4 \times 10^4$} 
& \multicolumn{3}{|c|}{$N=6 \times 10^4$} 
& \multicolumn{3}{|c|}{$N=6 \times 10^4$} 
& \multicolumn{3}{|c|}{$N=12 \times 10^4$} 
\\
\cline{2-16}
& \multicolumn{3}{|c|}{-} 
& \multicolumn{3}{|c|}{\emph{single-network}} 
& \multicolumn{3}{|c|}{\emph{single-network}} 
& \multicolumn{3}{|c|}{\emph{two-network}}
& \multicolumn{3}{|c|}{\emph{two-network}} 
\\
\cline{2-16}
&\multicolumn{3}{|c|}{\makecell[c]{$\epsilon_u\quad\, \epsilon_v\quad\, \epsilon_p$}} 
&\multicolumn{3}{|c|}{\makecell[c]{$\epsilon_u\quad\, \epsilon_v \quad\, \epsilon_p$}} 
&\multicolumn{3}{|c|}{\makecell[c]{$\epsilon_u\quad\, \epsilon_v \quad\, \epsilon_p$}}
&\multicolumn{3}{|c|}{\makecell[c]{$\epsilon_u\quad\, \epsilon_v \quad\, \epsilon_p$}}
&\multicolumn{3}{|c|}{\makecell[c]{$\epsilon_u\quad\, \epsilon_v \quad\, \epsilon_p$}}
\\
\midrule
\textcolor{blue}{\textbf{1}}
&\multicolumn{3}{|c|}{\makecell[c]{89.0\, 88.6\, 94.0}} 
&\multicolumn{3}{|c|}{\makecell[c]{3.3\, 3.4\, 6.9}}
&\multicolumn{3}{|c|}{\makecell[c]{4.2\, 4.3\, 8.6}} 
&\multicolumn{3}{|c|}{\makecell[c]{4.4\, 4.5\, 6.9}}
&\multicolumn{3}{|c|}{\makecell[c]{4.5\, 4.3\, 9.9}}
\\
\midrule
\textcolor{blue}{\textbf{2}}
&\multicolumn{3}{|c|}{\makecell[c]{89.2\, 88.4\, 98.6}} 
&\multicolumn{3}{|c|}{\makecell[c]{2.1\, 2.2\, 4.4}}
&\multicolumn{3}{|c|}{\makecell[c]{2.0\, 1.7\, 4.1}} 
&\multicolumn{3}{|c|}{\makecell[c]{3.2\, 3.3\, 6.7}} 
&\multicolumn{3}{|c|}{\makecell[c]{7.8\, 7.8\, 17.2}} 
\\
\midrule
\textcolor{blue}{\textbf{3}}
&\multicolumn{3}{|c|}{\makecell[c]{36.5\, 37.9\, 58.8}} 
&\multicolumn{3}{|c|}{\makecell[c]{2.8\, 2.9\, 5.8}}
&\multicolumn{3}{|c|}{\makecell[c]{4.1\, 4.2\, 8.5}} 
&\multicolumn{3}{|c|}{\makecell[c]{3.9\, 4.0\, 6.5}} 
&\multicolumn{3}{|c|}{\makecell[c]{8.3\, 8.4\, 18.5}}
\\
\midrule
\textcolor{blue}{\textbf{4}}
&\multicolumn{3}{|c|}{\makecell[c]{23.9\, 24.3\, 41.7}} 
&\multicolumn{3}{|c|}{\makecell[c]{3.7\, 3.8\, 7.6}}
&\multicolumn{3}{|c|}{\makecell[c]{4.7\, 4.8\, 9.6}} 
&\multicolumn{3}{|c|}{\makecell[c]{1.8\, 1.9\, 4.0}} 
&\multicolumn{3}{|c|}{\makecell[c]{7.4\, 7.6\, 16.6}}\\
\midrule
\textcolor{blue}{\textbf{5}}
&\multicolumn{3}{|c|}{\makecell[c]{21.9\, 22.7\, 39.5}} 
&\multicolumn{3}{|c|}{\makecell[c]{3.9\, 4.0\, 7.9}} 
&\multicolumn{3}{|c|}{\makecell[c]{4.5\, 4.5\, 9.1}}
&\multicolumn{3}{|c|}{\makecell[c]{2.4\, 2.5\, 5.2}}
&\multicolumn{3}{|c|}{\makecell[c]{6.3\, 6.7\, 13.9}}
\\
\midrule
\textcolor{red}{\textbf{mean}} 
&\multicolumn{3}{|c|}{\makecell[c]{\textcolor{red}{55.5\, 55.7\, 66.8}}}
&\multicolumn{3}{|c|}{\makecell[c]{\textcolor{red}
{3.6\, 3.7\, 7.4}}} 
&\multicolumn{3}{|c|}{\makecell[c]{\textcolor{red}{4.5\, 4.6\,	 9.6}}}
&\multicolumn{3}{|c|}{\makecell[c]{\textcolor{red}{3.1\, 3.2\,	 5.9}}}
&\multicolumn{3}{|c|}{\makecell[c]{\textcolor{red}{6.9\, 7.0\, 15.2}}} 
 \\
\bottomrule
\end{NiceTabular}
\end{adjustbox}
\end{center}
\label{rpe_variable_lr}
\end{table}
\renewcommand{\arraystretch}{1.0}
\begin{table}
\caption{Learning rate, training epochs and $\alpha$. Note that the entire training process can be divided into several stages; different $lr$ and $\alpha$ are used in different stage.}
\begin{center}
\begin{adjustbox}{width=0.95\textwidth}
\begin{tabular}{cccccccc}
\toprule
&Stage & 1 & 2 & 3 & 4 & 5 & 6\\
\midrule
\multirow{3}{8em}{\textcolor{blue}{$Re=2,000$ \emph{single-network}}} 
&Training epochs & $3 \times 10^{5}$ & $3 \times 10^{5}$ & $3 \times 10^{5}$& $5 \times 10^{5}$ & $10 \times 10^{5}$ & -\\

&Learning rate & $1 \times 10^{-3}$ & $2 \times 10^{-4}$ & $4 \times 10^{-5}$& $1 \times 10^{-5}$ & $2 \times 10^{-6}$ & - \\
&$\alpha$ & 0.05 & 0.03 & 0.01 & 0.005 & 0.002 & - \\
\midrule
\multirow{3}{8em}{\textcolor{blue}{$Re=3,000$ \\ \emph{single-network}}} 
&Training epochs & $3 \times 10^{5}$ & $3 \times 10^{5}$ & $3 \times 10^{5}$& $5 \times 10^{5}$ & $10 \times 10^{5}$ & -
\\

&Learning rate & $1 \times 10^{-3}$ & $2 \times 10^{-4}$ & $4 \times 10^{-5}$& $1 \times 10^{-5}$ & $2 \times 10^{-6}$ & -\\
&$\alpha$ & 0.05 & 0.03 & 0.01 & 0.005 &  0.002 & -\\
\midrule
\multirow{3}{8em}{\textcolor{red}{$Re=3,000$ \emph{two-network}}} 
&Training epochs & $2 \times 10^{5}$ & $2 \times 10^{5}$ & $2 \times 10^{5}$& $5 \times 10^{5}$ & $5 \times 10^{5}$ & $5 \times 10^{5}$\\

&Learning rate & $1 \times 10^{-3}$ & $2 \times 10^{-4}$ & $4 \times 10^{-5}$& $1 \times 10^{-5}$ & $2 \times 10^{-6}$ & $2 \times 10^{-6}$\\

&$\alpha$ & 0.05 &  0.03 & 0.01& 0.005 &  0.002 & 0.002\\
\midrule
\multirow{3}{8em}{\textcolor{red}{$Re=5,000$ \emph{two-network}}} 
&Training epochs & $5 \times 10^{5}$ & $5 \times 10^{5}$ & $5 \times 10^{5}$& $5 \times 10^{5}$ & $5 \times 10^{5}$ & $5 \times 10^{5}$\\

&Learning rate & $1 \times 10^{-3}$ & $2 \times 10^{-4}$ & $4 \times 10^{-5}$& $1 \times 10^{-5}$ & $2 \times 10^{-6}$ & $2 \times 10^{-6}$\\

&$\alpha$ & 0.05 &  0.03 & 0.01& 0.005 &  0.002 & 0.002\\
\bottomrule
\end{tabular}
\end{adjustbox}
\end{center}
\label{Re2K_lr}
\end{table}

\clearpage

\section*{Acknowledgments}
ZW would like to thank Dr. Shengze Cai at Zhejiang University for his help on NSFnet. ZW acknowledges supports by the Fundamental Research Funds for the Central Universities (DUT21RC(3)063).
GEK acknowledges support by the DOE SEA-CROGS project (DE-SC0023191) and the MURI-AFOSR FA9550-20-1-0358 project. XJ acknowledges supports by the Scientific Innovation Program from Dalian University of Technology under Grant (DUT22LAB502) and the Dalian Department of Science and Technology (2022RG10).

\bibliographystyle{unsrt}  
\bibliography{reference}

\end{document}